\documentclass{article}

\usepackage{graphicx}
\usepackage{multirow}
\usepackage{amsmath,amssymb,amsfonts}
\usepackage{amsthm}
\usepackage{mathrsfs}
\usepackage[title]{appendix}
\usepackage{xcolor}
\usepackage{textcomp}
\usepackage{manyfoot}
\usepackage{booktabs}
\usepackage{algorithm}
\usepackage{algorithmicx}
\usepackage{algpseudocode}
\usepackage{listings}

\usepackage{pgf,tikz}
\usetikzlibrary{arrows}
\usepackage{enumerate}
\usepackage{bbm}
\usepackage{color}
\usepackage{float}

\newtheorem{proposition}{Proposition}
\newtheorem{theorem}{Theorem}

\newtheorem{remark}{Remark}
\newtheorem{lemma}{Lemma}
\newtheorem{corollary}{Corollary}

\newcommand{\p}{\mathbbm{P}}
\newcommand{\e}{\mathbbm{E}}
\newcommand{\N}{\mathbbm{N}}

\newcommand{\LL}{\mathscr{L}}
\newcommand{\HH}{\mathcal{H}}
\newcommand{\M}{\mathcal{M}}
\newcommand{\OO}{\mathcal{O}}

\begin{document}

\title{Counting approximate breakpoint medians of random genomes}

\author{Poly H. da Silva, Arash Jamshidpey\footnote{Email: arash.jamshidpey@berkeley.edu}, David Sankoff}
\date{}

\maketitle

\begin{abstract}
Unsigned linear unichromosomal genomes with genes labeled $1,\ldots,n$
are represented by permutations of these labels, whose adjacencies are the unordered pairs of genes occurring consecutively. The breakpoint distance counts the adjacencies of one genome that are absent from the
other, and a breakpoint median minimizes the total breakpoint distance to a given set of genomes. Deciding whether a permutation is a median often
requires a minimization over the whole symmetric group, which makes the median set resistant to exact counting. We replace this global
condition by a local one: a $c$-approximate median is a genome having at most $c$ adjacencies absent from all the input genomes. For a fixed genome we introduce its \textit{approximate median inverse}, the set of
ordered $k$-tuples of genomes admitting it as a $c$-approximate median, and we construct and count this set explicitly. 
For $k$ genomes chosen uniformly and independently at random, this gives the probability that a given genome is a $c$-approximate median and the expected number of such approximate medians. The relaxation is not only technical: approximate medians and near-optimal genomes coincide up to a shift of the tolerance that remains bounded in probability as the number of genes grows. Finally, we derive an explicit upper bound on the probability that a given genome is an exact median, and on the expected number of exact breakpoint medians.

\vspace*{3mm}
\textbf{keyword:} Random permutation, breakpoint median, median inverse.
\end{abstract}

\section{Introduction}\label{section1}
A simple but effective way to measure gene-order dissimilarity between two genomes is to count their breakpoints. A breakpoint of a genome $G$ with respect to another genome $G'$ is a pair of genes that are adjacent in $G$ but not in $G'$. For genomes with the same gene content, the breakpoint distance is the number of such breakpoints.
Given a distance $d$ on a space of genomes, a median of genomes $G_1,\ldots,G_k$ is a genome minimizing the total distance function $\sum_{i=1}^{k} d(\cdot ,G_i)$. The breakpoint median problem, introduced in~\cite{sankoff97}, is the case in which $d$ is the breakpoint distance.

Medians are often employed to reconstruct gene-order evolutionary histories in the small phylogeny problem, in which ancestral genomes at the internal vertices of a given phylogenetic tree are inferred from genomes observed at its leaves. Under suitable evolutionary conditions, approximate medians,  genomes whose total distance to the input genomes is close to the minimum, can provide useful estimates of ancestral gene orders (see~\cite{jam13,jam16,jam17} for mathematical studies), while also capturing genetic similarities shared among the input genomes. Their reliability for ancestral inference, however, depends on the degree of evolutionary divergence, since the ancestral genome may cease to be an approximate median beyond an initial low-divergence regime~\cite{jam13,jam17,miardan23}.
Despite their importance, finding and constructing medians can be challenging~\cite{dSJS24,larlee14}. In fact, the median problem is NP-hard for several genomic distances, including the breakpoint distance on unsigned unichromosomal
genomes~\cite{bryant98,caprara03,tannier09,fertincombinatorics}.
Moreover, the breakpoint median need not be unique. Simulation studies
indicate that distinct median solutions may be widely separated and may tend
toward the input genomes as evolutionary divergence
among the input genomes increases~\cite{haghighi12}. Recent algorithmic work~\cite{dSJS25b} further illustrates this multiplicity and diversity through numerical examples in which approximate and exact breakpoint medians of triples of random genomes are enumerated and their distance profiles relative to the input genomes are examined. These observations motivate studying the size and structure of the
exact and approximate breakpoint median sets, both as a combinatorial object and in ancestral
genome inference.

Motivated by these questions, we formulate the problem in a combinatorial
and probabilistic setting. Labeling the genes by
$[n]:=\{1,\ldots,n\}$, we represent a linear unichromosomal genome of size
$n$ by a permutation of $[n]$. We denote by $S_n$ the set of all
permutations of $[n]$, or equivalently, the set of all such genomes with
gene content $[n]$.
Let $\xi_1^{(n)},\ldots,\xi_k^{(n)}$ be chosen uniformly and independently at random from $S_n$, and let $\mathcal{M}$ denote the set of their breakpoint medians. For a given set $A\subseteq S_n$, one may ask for the probability $\p(A\subseteq\mathcal{M})$. For sets $A$ containing more than one genome,
this probability depends on their relative configuration, including their mutual breakpoint distances, and the problem becomes considerably more involved. In this paper, we take a first step toward this
general problem by focusing on the singleton case $A=\{x\}$, that is, on the
probability that a given permutation $x$ belongs to $\mathcal{M}$.
Even the singleton case leads to a substantial combinatorial counting problem.
We introduce the \emph{median inverse} and obtain upper bounds for $\mathbb P(x\in \mathcal{M})$, $x\in S_n$, through a tractable relaxation based on approximate medians, which we also show to be the near-optimal solutions of the median problem.
By the left-invariance of the breakpoint distance and the uniform distribution of the input permutations, the singleton probability is the same for every $x\in S_n$. Consequently,
$\e|\mathcal{M}|=n!~\p(x\in\mathcal{M})$,
so upper bounds for the singleton probability also yield upper bounds
for the expected number of breakpoint medians.

For a fixed $k$, the median inverse of a permutation $\pi\in S_n$ is the set of
all ordered \mbox{$k$-tuples} $(y_1,\ldots,y_k)$ of permutations in $S_n$
for which $\pi$ is an exact breakpoint median. An exact median may contain adjacencies that are absent
from all the input genomes. We show, however, that the number of such
adjacencies admits an explicit upper bound and that, for random input
genomes, this bound stays bounded in probability, and is therefore negligible compared with the $n-1$ adjacencies of a genome. This motivates the notion of a
\emph{$c$-approximate breakpoint median}, defined as a genome having at most
$c$ adjacencies absent from all the inputs. For a suitably growing tolerance $c=c_n$,
the corresponding approximate median set contains every exact median with
high probability and therefore provides an explicitly countable outer
approximation of the exact median set. We also show that
approximate medians are of interest in their own right: they are
precisely the near-optimal solutions of the median problem, up to a
shift of the tolerance that is bounded in probability for random
genomes.
Similarly, for a fixed permutation $\pi$, we define its \emph{$c$-approximate median
inverse} as the set of all ordered $k$-tuples of genomes for which $\pi$ is
a $c$-approximate median. We construct and count this set and use it to determine the probability that a given genome is a $c$-approximate median of $k$ uniformly and independently chosen random genomes and to derive an explicit expression for the expected number of approximate medians.
For exact medians we derive an explicit upper bound on the probability
that a given genome is an exact median and on the expected number of
exact breakpoint medians, which in particular shows that this probability tends to zero faster than any exponential in $n$, at an explicit rate. Our framework also provides a constructive procedure for generating
approximate medians with prescribed breakpoint distances, together with a dynamic-programming procedure for exact evaluation of the counting formula, with complexity bound and numerical illustration.

The paper is organized as follows. Section~\ref{Sec-Def} introduces the
main definitions, including the median and segment sets, and
Section~\ref{brief-decsription} overviews the main results.
Section~\ref{Sec-Results} contains the principal theoretical and counting results. We first consider genomes with maximum or near-maximum pairwise breakpoint distances and count ordered $k$-tuples for which every adjacency of a given permutation occurs in at least one input genome. This special case motivates the general framework of Theorems~\ref{thmalpha}--\ref{seth}. Theorem~\ref{thmalpha} bounds the number of adjacencies of an exact median, absent from all inputs, and Theorem~\ref{thm-inclusion} gives the asymptotic behavior of this bound for random genomes. Theorem~\ref{thm:near-optimal} identifies approximate medians with the near-optimal solutions of the median problem. Theorems~\ref{thm:expected number} and~\ref{seth} give the probability that a fixed permutation is a $c$-approximate median of $k$ independent uniform random genomes, and the expected number of such approximate medians, and Theorem~\ref{thm:unconditional} gives explicit upper bounds for the corresponding quantities for exact medians. Finally, Theorem~\ref{thm:abundance} compares the sizes of the exact and approximate median sets for random genomes. Section~\ref{sec:conclusion} concludes the paper with a discussion of the results.

\section{Definitions}\label{Sec-Def}
We assume that there are no duplicated genes. Thus, unsigned linear unichromosomal genomes with $n$ genes or markers labeled by $1,\cdots,n$ are represented by permutations on $[n]:=\{1,\cdots,n\}$. Let $S_n$ denote the group of all permutations on $[n]$, endowed with  function composition as the multiplication operation, and denote by $id:=id^{(n)}$ the identity permutation $1 \ 2 \ 3 \ ... \ n$. For a permutation $\pi := \pi_1 \  ... \ \pi_n$, any unordered pair $\{\pi_i , \pi_{i+1}\}=\{\pi_{i+1} , \pi_i\}$, for $i=1, ..., n-1$, is called
a gene adjacency, or simply an adjacency, of $\pi$. The set of all adjacencies of $\pi$ is denoted by $\mathcal{A}_{\pi}$. Also, $\mathcal A_{x_1,...,x_k}$ denotes the set of all common adjacencies of $x_1,...,x_k\in S_n$. The \textit{breakpoint distance} (bp distance) between $x,y\in S_n$ is defined by $$d(x,y)=d^{(n)}(x,y):=n-1-|\mathcal A_{x,y}|=|\mathcal{A}_x\Delta\mathcal{A}_y|/2,$$
where $\Delta$ denotes the symmetric difference of sets.
The bp distance is a left-invariant pseudometric on $S_n$, that is, for any $x,y,z \in S_n$, $d(x,y)=d(zx, zy)$, where the product is the function composition $(zx)(i)=z(x(i))$. Indeed, $\mathcal A_{zx}=\{\{z(a),z(b)\}:\{a,b\}\in\mathcal A_x\}$, so left multiplication relabels the genes and preserves the number of common adjacencies.

Define the total breakpoint distance of $x\in S_n$ to a set of permutations $A\subseteq S_n$ by
$
d_T(x,A):=\sum_{y\in A} d(x,y).
$
A \emph{median} of $A\subseteq S_n$  is a permutation in $S_n$ (not necessarily unique) whose total distance to $A$ takes the minimum value, i.e.  it is a permutation $x\in S_n$ such that
$
d_T(x,A)=\min_{y\in S_n} d_T(y,A).
$
Furthermore, the median value of $A$, denoted by $\mu_n(A)$, is the minimum value of the total distance function to $A$. We denote by $\mathcal{M}_n(A)$ the set of all breakpoint medians of $A\subseteq S_n$. 
For $k\in\mathbbm N$, let $S_n^k$ denote the set of ordered $k$-tuples of permutations in $S_n$, with repetitions allowed. For $(x_1,\ldots,x_k)\in S_n^k$, its medians and median value are defined by
$\mathcal M_n(x_1,\ldots,x_k):=\mathcal M_n(\{x_1,\ldots,x_k\})$ and
$\mu_n(x_1,\ldots,x_k):=\mu_n(\{x_1,\ldots,x_k\})$. Note that $\mathcal{M}_n(A)\neq \emptyset$.
 
It is not hard to see that $z\in S_n$ is a median of $x,y\in S_n$ iff
$d(x,y)=d(x,z)+d(z,y)$. Such a $z$ is called a geodesic point of
$x,y$~\cite{jam14}, and the set of all geodesic points is denoted by
$\overline{[x,y]}$.

A segment of $S_n$ is a set of consecutive (gene) adjacencies of a permutation (genome) in $S_n$. More precisely, a segment of length $r\in [n-1]$ is a set of adjacencies 
$
\{\{n_0,n_1\},\{n_1,n_2\},...,\{n_{r-2},n_{r-1}\},\{n_{r-1},n_r\}\},
$
for $r+1$ distinct numbers (genes) $n_0,n_1,...,n_r\in [n]$. In particular, the empty set $\emptyset$ is regarded as a segment of length $0$. We denote by $|s|$ the length of a segment $s$. 
We say two segments $s$ and $s'$ are strongly disjoint if they share no genes (numbers). They are disjoint if they do not share any common adjacencies. Otherwise, we say they intersect each other. 

\begin{center}
\begin{figure}
\begin{center}
\includegraphics[scale=0.48, trim = 0.5cm 10.5cm 3cm 3cm,clip]{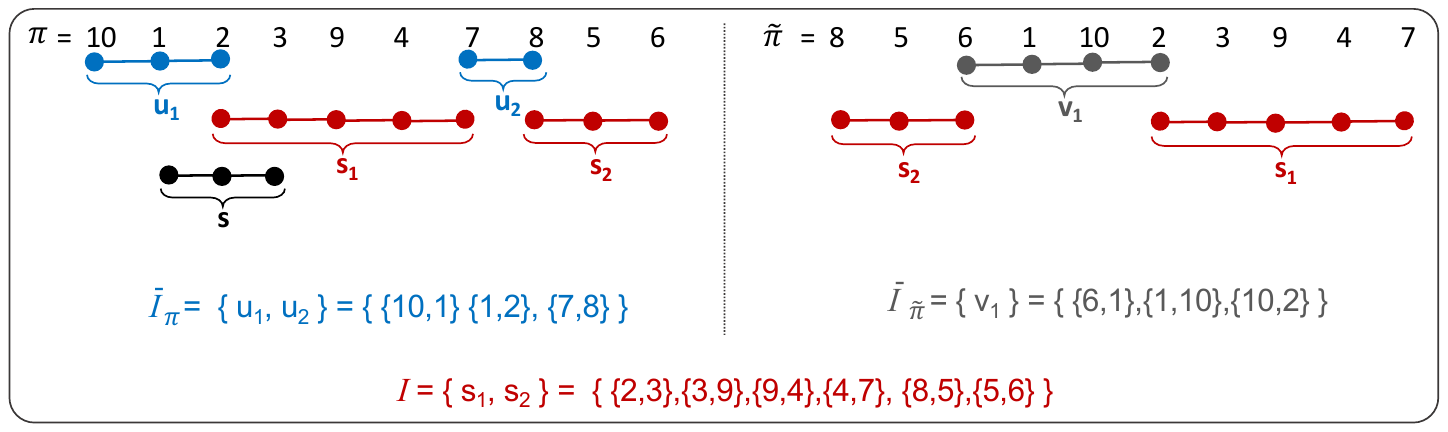}
\end{center}
\caption{The segment set $I$ is contained in permutations $\pi$ and $\tilde\pi$. 
As $\mathcal A_\pi\neq\mathcal A_{\tilde\pi}$,  the corresponding complement segment sets of \(I\) are different, i.e., $\bar I_{\pi}\neq\bar I_{\tilde\pi}$. 
The segments $s_1$ and $s_2$ are strongly disjoint,  $s_1$ and $u_1$ are disjoint, and $s_1$ and $s$ intersect.  The segments $s$ and $u_1$ are consistent with union $\{\{10,1\},\{1,2\},\{2,3\}\}$, while $s$ and $v_1$ are not consistent.
}\label{fig-seg}
\end{figure}
\end{center}

A segment set $I$ of a permutation $\pi\in S_n$ is a subset of the gene adjacencies of $\pi$, i.e. $I\subseteq\mathcal{A}_\pi$. Alternatively, we say $\pi$ contains $I$. 
Note that a segment set can be contained in more than one permutation. Denote by $\mathcal{I}^{(n)}$ the set of all segment sets of $S_n$. By a segment of a segment set $I$ we mean a maximal nonempty
segment contained in $I$. Also we denote by $\|I\|$ the number of segments of $I$. We emphasize that ``$|\ .\ |$'' is used for both cardinality of a set and absolute value of a real number. So for a segment set $I$ with segments $s_1,\cdots,s_{\|I\|}$, 
$
|I|=\sum_{i=1}^{\|I\|} |s_i|
$
counts the number of adjacencies of $I$. Two segment sets $I$ and $J$ (in particular two segments $s$ and $s'$, respectively) are said to be consistent, if their union is contained in $\mathcal{A}_\pi$, for some permutation $\pi$. 
We say segment sets $I_1,...,I_m$ complete each other if there exists a permutation $\pi$ such that $\cup_{i=1}^m I_i=\mathcal{A}_\pi$. In particular, the complement of a segment set $I$ with respect to $\pi\in S_n$ is the unique segment set $\Bar{I}_\pi:=\mathcal{A}_\pi\setminus I$; see Figure~\ref{fig-seg}.

\section{Brief description of results}\label{brief-decsription}

For $\pi\in S_n$, we define its
\emph{$k$-median inverse} by
\[
\M_{n,k}^{-1}(\pi)
:=
\left\{
(x_1,\ldots,x_k)\in S_n^k:
\pi\in\M_n(x_1,\ldots,x_k)
\right\}.
\]
More generally, for a set $A\subseteq S_n$, one may define its
$k$-median inverse as the set of all ordered $k$-tuples
$(x_1,\ldots,x_k)\in S_n^k$ satisfying
$A\subseteq\M_n(x_1,\ldots,x_k)$. The present paper considers only the
singleton case $|A|=1$.
If $\xi_1^{(n)},\ldots,\xi_k^{(n)}$ are chosen uniformly and
independently at random from $S_n$, then
\[
\mathbbm P\left(
\pi\in\M_n(\xi_1^{(n)},\ldots,\xi_k^{(n)})
\right)
=
\frac{|\M_{n,k}^{-1}(\pi)|}{(n!)^k}.
\]
By the left-invariance of the breakpoint distance, this probability is the same for every $\pi\in S_n$ and therefore determines the expected number of exact breakpoint medians.

Our first main result controls the adjacencies of an exact median that
are absent from all the input permutations. For
$X=\{x_1,\ldots,x_k\}\subseteq S_n$, Theorem~\ref{thmalpha} gives an
explicit upper bound $\mathcal O_n(X)$ for
$\max_{\pi\in\M_n(X)}
\left|
\mathcal A_\pi\setminus
\cup_{i=1}^k\mathcal A_{x_i}
\right|.$
For independent uniform random permutations
$\xi_1^{(n)},\ldots,\xi_k^{(n)}$, Theorem~\ref{thm-inclusion} shows
that, for every diverging sequence $(a_n)_{n\in\mathbbm N}$,
$\mathcal O_n(\{\xi_1^{(n)},\ldots,\xi_k^{(n)}\})/a_n
\to 0$
in probability as $n\to\infty$. In intuitive terms, this means that, for large random genomes, an exact median is unlikely to contain more than a small number of adjacencies that are absent from all the input genomes. In particular, by choosing a diverging sequence $(a_n)_{n\in\mathbbm{N}}$ satisfying $a_n=o(n)$, we see that this number is negligible compared with the total number $n-1$ of adjacencies in the median.
This motivates the following relaxation. For $c\geq0$, define the
\emph{$c$-approximate median set} of $X\subseteq S_n$ by
\begin{equation}\label{L(n,c)}
\mathscr L_{n,c}(X)
:=
\left\{
\pi\in S_n:
\left|
\mathcal A_\pi\setminus
\cup_{x\in X}\mathcal A_x
\right|
\leq c
\right\},
\end{equation}
and, for $\pi\in S_n$, define its
\emph{$c$-approximate median inverse} by
\begin{equation}\label{L-inv}
\mathscr L_{n,k,c}^{-1}(\pi)
:=
\left\{
(y_1,\ldots,y_k)\in S_n^k:
\left|
\mathcal A_\pi\setminus
\cup_{i=1}^k\mathcal A_{y_i}
\right|
\leq c
\right\}.
\end{equation}
When $X=\{x_1,\ldots,x_k\}$, we also write
$\mathscr L_{n,c}(x_1,\ldots,x_k):=\mathscr L_{n,c}(X)$.

By Theorem~\ref{thmalpha}, on the event
$\{\mathcal O_n(\xi_1^{(n)},\ldots,\xi_k^{(n)})\leq a_n\}$ every exact
median is an $a_n$-approximate median, and Theorem~\ref{thm-inclusion}
shows that this event has probability tending to one whenever
$a_n\to\infty$. Counting the approximate median inverse therefore
controls the exact median set outside an event of vanishing
probability; see~\eqref{eq:restricted}.

Approximate medians are also of interest in their own right. Writing
$\M_n^{c}(X)$ for the set of permutations whose total distance to $X$
exceeds the median value by at most $c$,
Theorem~\ref{thm:near-optimal} shows that
$\LL_{n,c}(X)\subseteq\M_n^{\,c+\Lambda_n(X)}(X)$ and, conversely,
$\M_n^{c}(X)\subseteq\LL_{n,\mathcal O_n(X)+c}(X)$, where $\Lambda_n(X)$
and $\mathcal O_n(X)$ measure the overlaps among the adjacency sets of
the input genomes. Thus approximate medians and near-optimal
permutations coincide up to a shift of the tolerance, and by
Corollary~\ref{cor:sandwich} this shift is bounded in probability for
random genomes. Taking $c=0$ in the second inclusion gives
$\M_n(X)\subseteq\LL_{n,\mathcal O_n(X)}(X)$ for every set of input
genomes.

For fixed $c\geq0$, the approximate quantities can be computed exactly.
Theorem~\ref{thm:expected number} gives
\[
\e\left|
\mathscr L_{n,c}
(\xi_1^{(n)},\ldots,\xi_k^{(n)})
\right|
=
\frac{
|\mathscr L_{n,k,c}^{-1}(\pi)|
}{(n!)^{k-1}},
\]
and Theorem~\ref{seth} provides the combinatorial ingredients needed to
evaluate\break $|\mathscr L_{n,k,c}^{-1}(\pi)|$. For exact medians,
Theorem~\ref{thm:unconditional} gives an explicit upper bound for the
probability that a given permutation is an exact median and for the
expected number of exact medians, valid without any restriction to a
favorable event, and shows in particular that this probability tends
to zero, faster than any exponential in $n$. Finally, Theorem~\ref{thm:abundance} compares the exact and approximate median sets of random genomes: $|\M_n|/|\LL_{n,\mathcal O_n+s}|=O_{\p}(n^{-s})$ for every fixed integer $s\geq1$. Our framework also gives a constructive procedure for generating
approximate medians with prescribed breakpoint distances, and a
dynamic-programming procedure for exact evaluation of the counting
formula, whose complexity we analyze and illustrate numerically.

\section{Results}\label{Sec-Results}
For $n\geq 2$ and a permutation $x\in S_n$, it is clear that $\mathcal{M}_{n,1}^{-1}(x)=\{x,y\}$, where $y\neq x$ is the unique permutation for which $d(x,y)=0$. For instance, $\mathcal{M}_{n,1}^{-1}(id)=\{id,y\}$, where $y=n \ n-1 \ \cdots \ 2 \ 1$.  Also,
$
\mathcal{M}_{n,2}^{-1}(x)=\{(x_1,x_2)\in S_n^2: \ x\in\overline{[x_1,x_2]}\}.
$
It is well known that the expected number of common adjacencies of two random permutations is $O(1)$; see Proposition~\ref{tools}. This means that, with high probability, the random permutations are at approximately maximum distance from each other. Hence, to study the median of $k$ random permutations, we need to review some results for the median of $k$ permutations with pairwise maximum distance $n-1$ from each other. In particular, recall that a permutation $\pi$ is a median of $k$ permutations with pairwise maximum distance $n-1$ from each other, if and only if every adjacency of $\pi$ is an adjacency of at least one of those $k$ permutations. In other words we have the following proposition.
\begin{proposition}[\cite{jam14}]\label{extreme}
Let $\emptyset\neq X \subseteq S_n$ such that $d(x,y)=n-1$ for any distinct $x,y\in X$. Then $\pi$ is a median of $X$ if and only if $\mathcal A_{\pi} \subseteq \cup_{x\in X} \mathcal A_x$.
\end{proposition}
This is not true in general when the distances of input genomes are not the maximum value $n-1$. For example, if $n=9$, $x=2 \ 7 \ 5 \ 6 \ 8 \ 3 \ 9 \ 4 \ 1$ and $\pi=6 \ 8 \ 9 \ 3 \ 4 \ 1 \ 2 \ 7 \ 5$, then every adjacency of $\pi$ is either an adjacency of $id=id^{(n)}$ or an adjacency of $x$,
but $d^{(9)}(id,x)=7<d^{(9)}(id,\pi)+d^{(9)}(\pi,x)=8$. In fact, this happens because all common adjacencies of $id$ and $x$ must be adjacencies of $\pi$ in order to have  $\pi \in \overline{[id,x]}$ as stated in the next proposition.

\begin{proposition}[\cite{jam14}]\label{dist}
Let $x,y \in S_n$. Then $z \in \overline{[x,y]}$ if and only if $\mathcal A_{x,y} \subseteq \mathcal A_z \subseteq \mathcal A_x \cup \mathcal A_y$.
\end{proposition}
Proposition~\ref{extreme} implies that, for any
$X\subseteq S_n$ whose elements have pairwise breakpoint distance $n-1$,
$\M_n(X)=\LL_{n,0}(X).$
Let
\[
V_{n,k}
:=
\left\{
(x_1,\ldots,x_k)\in S_n^k:
d(x_i,x_j)=n-1 \text{ for all } i\neq j
\right\}.
\]
Then, for every $\pi\in S_n$,
$\LL_{n,k,0}^{-1}(\pi)\cap V_{n,k}
=
\M_{n,k}^{-1}(\pi)\cap V_{n,k}.$
Thus, for input permutations at pairwise maximum distance, the
$0$-approximate median inverse coincides with the exact median inverse.
This provides a natural starting point for our counting framework. We first
study $\LL_{n,k,0}^{-1}(\pi)$, whose decomposition introduces the main
combinatorial objects used later for general values of $c$. The following
proposition gives this decomposition according to the adjacencies that each
input permutation shares with $\pi$.

\begin{proposition}\label{prop:L-inverse-decomposition}
Let $n,k\in\mathbbm N$ and $\pi\in S_n$. For a segment set
$I\in\mathcal I^{(n)}$ contained in $\pi$, define
$\mathcal H_\pi^{(n)}(I)
:=
\{x\in S_n:\mathcal A_{\pi,x}=I\},$
and let
\begin{equation}\label{index-partition}
\mathscr P_{k,0}^{(n)}(\pi)
:=
\left\{
(J_1,\ldots,J_k)\in(\mathcal I^{(n)})^k:
\cup_{i=1}^k J_i=\mathcal A_\pi
\right\}.
\end{equation}
Then
\[
\mathscr L_{n,k,0}^{-1}(\pi)
=
\bigsqcup_{(\tilde J_1,\ldots,\tilde J_k)
\in \mathscr P_{k,0}^{(n)}(\pi)}
\mathcal H_\pi^{(n)}(\tilde J_1)
\times\cdots\times
\mathcal H_\pi^{(n)}(\tilde J_k),
\]
and consequently,
\[
\left|\mathscr L_{n,k,0}^{-1}(\pi)\right|
=
\sum_{(\tilde J_1,\ldots,\tilde J_k)
\in \mathscr P_{k,0}^{(n)}(\pi)}
\prod_{i=1}^k
\left|\mathcal H_\pi^{(n)}(\tilde J_i)\right|.
\]
\end{proposition}

\begin{proof}
Observe first that, for permutations $x,y\in S_n$ and a segment set
$I$, we have $x\in\HH_y^{(n)}(I)$ if and only if 
$\mathcal A_{x,y}=I$, which is also equivalent to 
$y\in\HH_x^{(n)}(I)$.
Moreover, if $I$ and $I'$ are two distinct segment sets of
$\pi\in S_n$, then
$\HH_\pi^{(n)}(I)\cap\HH_\pi^{(n)}(I')=\emptyset.$
Indeed, if some $x\in S_n$ belonged to this intersection, then
$\mathcal A_{\pi,x}=I$ and $\mathcal A_{\pi,x}=I',$
which would imply $I=I'$, a contradiction.

Now, note that the segment sets $J_1,\ldots,J_k$ in the definition of
$\mathscr P_{k,0}^{(n)}(\pi)$ are not required to be disjoint and may therefore intersect one another. Let
$\mathcal J=(J_1,\ldots,J_k)$ and
$\mathcal J'=(J_1',\ldots,J_k')$
be two distinct elements of $\mathscr P_{k,0}^{(n)}(\pi)$. Since
$\mathcal J\neq\mathcal J'$, there exists
$r\in\{1,\ldots,k\}$ such that $J_r\neq J_r'$. It follows that
$\HH_\pi^{(n)}(J_r)\cap\HH_\pi^{(n)}(J_r')=\emptyset.$
Consequently,
\[
\left(
\HH_\pi^{(n)}(J_1)\times\cdots\times\HH_\pi^{(n)}(J_k)
\right)
\cap
\left(
\HH_\pi^{(n)}(J_1')\times\cdots\times\HH_\pi^{(n)}(J_k')
\right)
=
\emptyset.
\]
Thus, the union in the decomposition of
$\LL_{n,k,0}^{-1}(\pi)$ is disjoint.

Now let
$(x_1,\ldots,x_k)\in\mathscr L_{n,k,0}^{-1}(\pi)$. Then
$\mathcal A_\pi\subseteq\cup_{i=1}^k\mathcal A_{x_i}.$
For each $i$, set
$J_i:=\mathcal A_{\pi,x_i}.$
Since
\[
\bigcup_{i=1}^kJ_i
=
\mathcal A_\pi\cap\bigcup_{i=1}^k\mathcal A_{x_i}
=
\mathcal A_\pi,
\]
we have
$(J_1,\ldots,J_k)\in \mathscr P_{k,0}^{(n)}(\pi)$ and
$x_i\in \mathcal H_\pi^{(n)}(J_i)$ for every $i$. Thus,
\[
(x_1,\ldots,x_k)
\in
\mathcal H_\pi^{(n)}(J_1)\times\cdots\times \mathcal H_\pi^{(n)}(J_k).
\]

Conversely, suppose that
\[
(x_1,...,x_k)\in \bigcup\limits_{(\tilde{J}_1,...,\tilde{J}_k)\in \mathscr{P}_{k,0}^{(n)}(\pi)} \HH_{\pi}^{(n)}(\tilde J_1)\times ...\times \HH_{\pi}^{(n)}(\tilde J_k).
\]
Then, there exists $(J_1,...,J_k)\in \mathscr{P}_{k,0}^{(n)}(\pi)$ such that $x_i\in \HH_{\pi}^{(n)}(J_i)$ and $\mathcal{A}_{\pi,x_i}=J_i$, for $i=1,...,k$. Hence
\[
\mathcal A_\pi
=
\bigcup_{i=1}^kJ_i
=
\bigcup_{i=1}^k\mathcal A_{\pi,x_i}
\subseteq
\bigcup_{i=1}^k\mathcal A_{x_i}.
\]
Therefore,
$(x_1,\ldots,x_k)\in\mathscr L_{n,k,0}^{-1}(\pi)$.
The cardinality formula follows from the disjoint-union decomposition and
the cardinality of a Cartesian product.
\end{proof}

Thus, determining the cardinalities of the sets $\HH_{\pi}^{(n)}(\tilde J_i)$ plays an important role in counting the elements of $\mathscr{L}_{n,k,0}^{-1}(\pi)$.

We now consider the case in which the input permutations are chosen uniformly and independently at random from $S_n$. As $n$ becomes large, this setting resembles the case of permutations at pairwise maximum breakpoint distance, since two independent uniform random permutations typically share only a small number of adjacencies. The following result makes this observation precise. Parts (i) and (ii) give the exact mean and variance; classically, the number of common adjacencies of a uniform random permutation with $id$ converges in distribution to a Poisson random variable with mean $2$~\cite{wolfowitz44,kaplansky45}; see also~\cite{xu08} for other genome models.

\begin{proposition}\label{tools}
Let $\xi=\xi^{(n)}$ be a permutation chosen uniformly at random from $S_n$, and let $(a_n)_{n\in \N}$ be a sequence of real numbers tending to $\infty$
. Then 
\begin{enumerate}[i)]
\item $\e [d(id,\xi)]=n-1 -\frac{2(n-1)}{n}, \ n\in \N$.
\item $var(d(id,\xi))= (2-\frac{2}{n})(-1+\frac{2}{n})+\frac{4(n-2)^2}{n(n-1)},  \ n\ge 2$.
\item For any $\varepsilon >0$, 
$
\p(n-1-d(id,\xi)\geq \varepsilon a_n)\rightarrow 0, \text{ as } n\rightarrow \infty.
$
\end{enumerate}
\end{proposition}

\begin{proof}
It suffices to compute the mean and variance of $\zeta:=|\mathcal A_{id,\xi}|$. For an adjacency
$e=\{j,j+1\}\in\mathcal A_{id}$, the permutations containing $e$ are
obtained by treating $j,j+1$ as a single block in either of its two
orientations, so their number is $2(n-1)!$ and
$\p(e\in\mathcal A_\xi)=2/n$. Since $|\mathcal A_{id}|=n-1$, \[\e\,\zeta=\sum_{e\in\mathcal A_{id}}\p(e\in\mathcal A_\xi)={2(n-1)}/{n},\] which gives (i). For distinct $e,e'\in\mathcal A_{id}$, the same argument
applied to the blocks they form gives
\[
\p(e,e'\in\mathcal A_\xi)=
\begin{cases}
2(n-2)!/n!=\dfrac{2}{n(n-1)}, & \text{if $e$ and $e'$ share a gene},\\[2mm]
4(n-2)!/n!=\dfrac{4}{n(n-1)}, & \text{otherwise},
\end{cases}
\]
since in the first case $e\cup e'$ consists of three consecutive genes,
forming one block with two orientations, and in the second case $e$ and
$e'$ form two blocks, each with two orientations. There are $n-2$
unordered pairs of the first kind and
$\binom{n-1}{2}-(n-2)=(n-2)(n-3)/2$ of the second, whence
\[
\e\bigl[\zeta(\zeta-1)\bigr]
=2\sum_{\{e,e'\}}\p(e,e'\in\mathcal A_\xi)
=\frac{4(n-2)+4(n-2)(n-3)}{n(n-1)}
=\frac{4(n-2)^2}{n(n-1)},
\]
and $var(\zeta)=\e[\zeta(\zeta-1)]+\e\,\zeta-(\e\,\zeta)^2$ gives (ii).
Finally, (iii) follows from (i) and Markov's inequality, since
$\p(\zeta\geq\varepsilon a_n)\leq 2(n-1)/(n\varepsilon a_n)\to0$ as
$a_n\to\infty$.
\end{proof}
As mentioned above, when the permutations in $X\subseteq S_n$ are at pairwise maximum distance $n-1$, a permutation $\pi$ is a median of $X$ if and only if each of its adjacencies occurs in at least one permutation in $X$. This characterization does not hold for a general set
$X\subseteq S_n$. In particular, a median need not take all of its
adjacencies from
$\cup_{x\in X}\mathcal A_x.$
It is therefore natural to ask whether one can find a uniform upper bound on
$|
\mathcal A_\pi\setminus\cup_{x\in X}\mathcal A_x
|$
over all medians $\pi$ of $X$. The next theorem provides such a bound.
We first introduce the notation needed for its statement.

For distinct $x_1,\ldots,x_k\in S_n$, let $X=\{x_1,\ldots,x_k\}\subseteq S_n$, and define
$\mathcal B_X^X
=
\mathcal B_{x_1,\ldots,x_k}^X
:=
\mathcal A_{x_1,\ldots,x_k}.$
Let
$\mathcal B_{x_1,\ldots,x_{j-1},x_{j+1},\ldots,x_k}^X
:=
\mathcal A_{x_1,\ldots,x_{j-1},x_{j+1},\ldots,x_k}
\setminus
\mathcal B_{x_1,\ldots,x_k}^X,$ for each $j\in\{1,\ldots,k\}$. More generally, for
$\emptyset\neq U=\{x_{i_1},\ldots,x_{i_r}\}\subseteq X$, set
\[
\mathcal B_U^X
=
\mathcal B_{x_{i_1},\ldots,x_{i_r}}^X
:=
\mathcal A_U
\setminus
\bigcup_{U\subsetneqq V\subseteq X}\mathcal B_V^X.
\]
In other words, $\mathcal{B}_U^X$ includes all adjacencies that are common in every $x\in U$, but missing from every $y\in X\setminus U$.
Accordingly, for an adjacency $e$, let
$\nu_X(e):=|\{x\in X:\ e\in\mathcal A_x\}|$
denote its multiplicity, that is, the number of input permutations
containing $e$. Thus $\nu_X(e)=|U|$ for every $e\in\mathcal B_U^X$, and
$\nu_X(e)=0$ if $e\notin\cup_{x\in X}\mathcal A_x$. For $y\in S_n$ and $0\leq r\leq k$, define
\[
\eta_r(y)
:=
\left|\left\{e\in\mathcal A_y:\ \nu_X(e)=r\right\}\right|.\]
Thus, for $1\le r\le k$, $\eta_r(y)$ is the number of adjacencies of $y$ that occur in
exactly $r$ of the input permutations, while $\eta_0(y)$ is the number
of adjacencies of $y$ that occur in none of them. We also define
\begin{equation}\label{big_lambda-Def}
\Lambda_n(X)
:=
\sum_{r=2}^k(r-1)
\sum_{\substack{U\subseteq X\\ |U|=r}}
\left|\mathcal B_U^X\right|
=
\sum_{e\in\cup_{x\in X}\mathcal A_x}\bigl(\nu_X(e)-1\bigr).
\end{equation}
The quantity $\Lambda_n(X)$ counts all adjacencies that occur in at
least two input permutations, an adjacency occurring in exactly $r$
inputs being counted with weight $r-1$. For $j\in\{1,\ldots,k\}$ and $2\leq r\leq k-1$, let
\[
\beta_r^{(j)}(X)= \beta_r^{(j)}
:=
\sum_{\substack{
U\subseteq X\setminus\{x_j\}\\
|U|=r
}}
\left|\mathcal B_U^X\right|.
\]
Thus, $\beta_r^{(j)}(X)$ is the number of adjacencies occurring in
exactly $r$ input permutations, none of which is $x_j$. Also define
\[
\mathcal O_n^{(j)}(X)
:=
\sum_{r=2}^{k-1}(r-1)\beta_r^{(j)}(X), \quad \quad \mathcal O_n(X):=\min_{1\leq j\leq k}\mathcal O_n^{(j)}(X),
\]
with the convention that an empty sum equals zero. In particular,
$\mathcal O_n^{(j)}(X)=0$ when $k\le 2$. Thus $\mathcal O_n^{(j)}(X)$
collects the terms of $\Lambda_n(X)$ coming from adjacencies not
contained in $x_j$, and in particular
$\mathcal O_n^{(j)}(X)\le\Lambda_n(X)$ for every $j$. For $y_1,\ldots,y_k\in S_n$, we write $\mathcal O_n(y_1,\ldots,y_k):=\mathcal O_n(\{y_1,\ldots,y_k\})$. 

\begin{theorem}\label{thmalpha}
Let $X=\{x_1,\ldots,x_k\}\subseteq S_n$, where $x_1,\ldots,x_k$ are distinct and $k\geq2$. Then, for every $m\in\M_n(X)$ and every $j\in\{1,\ldots,k\}$,
\[
\left|
\mathcal A_m
\setminus
\bigcup_{x\in X}\mathcal A_x
\right|
\leq
\sum_{r=2}^k
(r-1)\bigl(\eta_r(m)-\eta_r(x_j)\bigr)
\leq
\mathcal O_n^{(j)}(X).
\]
Moreover, for any $i,j\in\{1,\ldots,k\}$,
$\mathcal O_n^{(j)}(X)-\mathcal O_n^{(i)}(X)
=
d_T(x_j,X)-d_T(x_i,X).$

\end{theorem}
\begin{remark}
It follows from Theorem~\ref{thmalpha} that 
$\left| \mathcal A_m \setminus \cup_{x\in X}\mathcal A_x \right| \leq \mathcal O_n(X)$
for every $m\in\M_n(X)$. Moreover, the identity 
$\mathcal O_n^{(j)}(X)-\mathcal O_n^{(i)}(X) = d_T(x_j,X)-d_T(x_i,X)$ 
shows that if $d_T(x_j,X) = \min_{1\leq i\leq k}d_T(x_i,X),$ then 
$\mathcal O_n(X)=\mathcal O_n^{(j)}(X).$
In particular, if the minimum total distance is attained by more than one input permutation, all such minimizers give the same value of $\mathcal O_n(X)$. When $k=2$, the sum defining $\mathcal O_n^{(j)}(X)$ is empty, and hence $\mathcal O_n(X)=0$; thus every median of $\{x_1,x_2\}$ takes all of its adjacencies from $\mathcal A_{x_1}\cup\mathcal A_{x_2}$, in agreement with Proposition~\ref{dist}. When $k=3$, 
$\mathcal O_n^{(j)}(X) = \beta_2^{(j)} =| \mathcal B_{X\setminus\{x_j\}}^X |,$ for $j=1,2,3.$
\end{remark}
\begin{proof}[Proof of Theorem~\ref{thmalpha}]
The quantities
$\eta_0(y),\eta_1(y),\ldots,\eta_k(y)$
partition the $n-1$ adjacencies of any permutation $y\in S_n$ according
to the number of input permutations in which they occur. Therefore,
\begin{equation}\label{eq:eta}
 \sum_{r=0}^k\eta_r(y)=n-1.   
\end{equation}
Moreover, an adjacency counted by $\eta_r(y)$ is shared by $y$ with
exactly $r$ input permutations. Hence,
\[
d_T(y,X)
=
\sum_{j=1}^k d(y,x_j)=k(n-1)-\sum_{r=1}^k r\eta_r(y).
\]

Fix $j\in\{1,\ldots,k\}$. Since $m$ is a median of $X$,
$d_T(m,X)\leq d_T(x_j,X),$
and therefore
\[
\sum_{r=1}^k r\eta_r(m)
\geq
\sum_{r=1}^k r\eta_r(x_j).
\]
On the other hand, from (\ref{eq:eta}), we have 
$\sum_{r=0}^k
\bigl(\eta_r(m)-\eta_r(x_j)\bigr)
=
0.$
Subtracting this equality from the preceding inequality gives
\[
0
\leq
\sum_{r=0}^k
(r-1)\bigl(\eta_r(m)-\eta_r(x_j)\bigr).
\]
Since every adjacency of $x_j$ occurs in at least one input permutation,
namely $x_j$ itself, we have
$\eta_0(x_j)=0.$
Furthermore, the coefficient of the $r=1$ term is zero. Thus,
\[
\eta_0(m)
\leq
\sum_{r=2}^k
(r-1)\bigl(\eta_r(m)-\eta_r(x_j)\bigr).
\]

It remains to bound
$\eta_r(m)-\eta_r(x_j)$. Fix $r\in\{2,\ldots,k-1\}$.
An adjacency can contribute positively to this difference only if it
belongs to $m$, occurs in exactly $r$ input permutations, and does not
occur in $x_j$. All such adjacencies are counted by
$\beta_r^{(j)}$.
By contrast, an adjacency occurring in $x_j$ either belongs to both
$m$ and $x_j$, in which case it cancels in the difference, or belongs
to $x_j$ but not to $m$, in which case it contributes negatively.
Therefore,
\[
\eta_r(m)-\eta_r(x_j)
\leq
\beta_r^{(j)},
\qquad 2\leq r\leq k-1.
\]

When $r=k$, every adjacency occurring in exactly $k$ input permutations
necessarily occurs in $x_j$. Hence, there is no positive contribution,
and
${\eta_k(m)-\eta_k(x_j)}$ $ \leq0.$
It follows that
\[
\eta_0(m)
\leq
\sum_{r=2}^{k-1}
(r-1)\beta_r^{(j)}
=
\mathcal O_n^{(j)}(X).
\]

To prove the relationship between
$\mathcal O_n^{(j)}(X)$ and $d_T(x_j,X)$, recall
from~\eqref{big_lambda-Def} and the definition of
$\mathcal O_n^{(j)}(X)$ that, for a fixed $j$, the terms of
$\Lambda_n(X)$ with $x_j\notin U$ add up to $\mathcal O_n^{(j)}(X)$.
The remaining terms are those for which $x_j\in U$. Every adjacency in such a set
$\mathcal B_U^X$ is shared by $x_j$ with exactly $|U|-1$ other input
permutations. Consequently,
$\Lambda_n(X)-\mathcal O_n^{(j)}(X)
=
\sum_{q\neq j}
|\mathcal A_{x_j,x_q}|.$
Since
$d_T(x_j,X)
=
(k-1)(n-1)
-
\sum_{q\neq j}
\left|
\mathcal A_{x_j,x_q}
\right|,$
we obtain
\[
\mathcal O_n^{(j)}(X)
=
\Lambda_n(X)-(k-1)(n-1)+d_T(x_j,X).
\]
Therefore, for any $i,j\in\{1,\ldots,k\}$,
$\mathcal O_n^{(j)}(X)-\mathcal O_n^{(i)}(X)
=
d_T(x_j,X)-d_T(x_i,X).$
\end{proof}
The key feature of Theorem~\ref{thmalpha} is that adjacencies occurring in exactly one input permutation do not appear in
the upper bound: $\mathcal O_n^{(j)}(X)$ is a weighted count of the adjacencies shared by at least two input permutations but absent from $x_j$, an adjacency occurring in exactly $r$ inputs having weight $r-1$. These weights come from the distance comparison in the proof, which gives, for every $y\in S_n$,
\begin{equation}\label{eq:dT-eta}
d_T(y,X)=(k-1)(n-1)+\eta_0(y)-\sum_{r=2}^k(r-1)\eta_r(y).
\end{equation}
Applying~\eqref{eq:dT-eta} to $y$ and $x_j$, with
$\eta_0(x_j)=0$, shows that the first inequality in
Theorem~\ref{thmalpha}, with $y$ in place of $m$, is equivalent to
$d_T(y,X)\leq d_T(x_j,X)$. Thus, the bounds of Theorem~\ref{thmalpha} hold for every
$y\in S_n$ satisfying
$d_T(y,X)\leq\min_{1\leq j\leq k}d_T(x_j,X)$.
The second inequality removes the dependence on $y$.

The smallness of $\mathcal O_n(X)$ is not restricted to random inputs. For fixed $k$, $\mathcal O_n(X)=o(n)$ whenever, after removing an input permutation minimizing the total distance to $X$, only $o(n)$ adjacencies are shared by two or more of the remaining input permutations. For independent uniformly distributed random permutations, these shared-adjacency counts are bounded in probability by Proposition~\ref{tools}; this is used in Theorem~\ref{thm-inclusion} below.

\begin{remark}\label{remark:upperb}
If the input permutations are at pairwise maximum breakpoint distance, then $\mathcal O_n(X)=0$, and Theorem~\ref{thmalpha} shows, in agreement with Proposition~\ref{extreme}, that every median takes all of its adjacencies from $\cup_{x\in X}\mathcal A_x$. The same conclusion holds whenever $\mathcal O_n(X)=0$, which can happen beyond the pairwise-maximum-distance setting. For example, let $x_1=id^{(6)}=1\ 2\ 3\ 4\ 5\ 6$, $x_2=4\ 6\ 5\ 1\ 3\ 2$, $x_3=4\ 2\ 6\ 5\ 1\ 3$ and $X=\{x_1,x_2,x_3\}$. Then $\mathcal A_{x_1,x_2}=\{\{2,3\},\{5,6\}\}$, $\mathcal A_{x_1,x_3}=\{\{5,6\}\}$, $\mathcal A_{x_2,x_3}=\{\{5,6\},\{1,5\},\{1,3\}\}$ and $\mathcal A_X=\{\{5,6\}\}$. Moreover, $d_T(x_1,X)=7$, $d_T(x_2,X)=5$ and $d_T(x_3,X)=6$, so $\mathcal O_6(X)=\mathcal O_6^{(2)}(X)=|\mathcal B_{x_1,x_3}^{X}|=|\mathcal A_{x_1,x_3}\setminus\mathcal A_X|=0$.
\end{remark}
\begin{remark}\label{rem:attained}
The bound of Theorem~\ref{thmalpha} can be attained even when
$\mathcal O_n(X)>0$; such examples can be found using the algorithms
in~\cite{dSJS25b}. Let
$x_1=1\ 2\ 3\ 4\ 5\ 6$,
$x_2=1\ 2\ 4\ 5\ 3\ 6$,
$x_3=1\ 4\ 5\ 3\ 2\ 6$,
and $X=\{x_1,x_2,x_3\}$. The adjacencies of multiplicity at least two
are $\{4,5\}$, with $\nu_X(\{4,5\})=3$, and $\{1,2\},\{2,3\},\{3,5\}$,
each of multiplicity two and absent from $x_3$, $x_2$ and $x_1$
respectively; hence $\mathcal O_6^{(j)}(X)=1$ for every $j$, and
$\mathcal O_6(X)=1$. Now let $m=1\ 2\ 3\ 5\ 4\ 6$, so that
$\mathcal A_m=\bigl\{\{1,2\},\{2,3\},\{3,5\},\{4,5\},\{4,6\}\bigr\}$.
Since $\{4,6\}$ occurs in none of the input permutations,
$\eta_0(m)=1=\mathcal O_6(X)$. By the identity established in the proof
of Theorem~\ref{thmalpha},
$d_T(y,X)=15-\sum_{e\in\mathcal A_y}\nu_X(e)$ for every $y\in S_6$, so
$d_T(m,X)=6$. A permutation $y$ with $d_T(y,X)\leq5$ would have to
contain all four adjacencies of multiplicity at least two together with
one further adjacency of positive multiplicity. These four adjacencies
form the path $1-2-3-5-4$, and no adjacency of positive multiplicity
joins the remaining gene $6$ to either endpoint. Hence $d_T(y,X)\geq6$
for every $y\in S_6$, so $m\in\M_6(X)$ and the bound is attained.
\end{remark}
Recall that $\mathscr L_{n,c}(X)$ denotes the set of
$c$-approximate medians of $X$. Theorem~\ref{thmalpha} shows that every
exact median is a $c$-approximate median whenever
$c\geq\mathcal O_n(X)$. Equivalently,
$\M_n(X)\subseteq\mathscr L_{n,c}(X),$ for $c\geq\mathcal O_n(X).$
In particular,
$\M_n(X)
\subseteq
\mathscr L_{n,\mathcal O_n(X)}(X).$
Thus, when $\mathcal O_n(X)$ is small,
$\mathscr L_{n,\mathcal O_n(X)}(X)$ provides a controlled relaxation of
the exact median set: it contains every exact median while allowing only
a limited number of adjacencies outside the union of the input adjacency
sets. Unlike the exact median
condition, which requires minimizing the total distance over all
permutations in $S_n$, membership in $\mathscr L_{n,c}(X)$ is determined
directly by the number of adjacencies of the candidate permutation that
do not appear in any of the inputs; it can therefore be studied through
the approximate median inverse introduced in
\eqref{L(n,c)} and~\eqref{L-inv}.

Recall also that the left-invariance of the breakpoint distance implies
$|\mathcal A_{x,y}|
=
|\mathcal A_{\pi x,\pi y}|$, for $\pi,x,y\in S_n.$
Therefore, for any $\pi,x\in S_n$ and
$X=\{x_1,\ldots,x_k\}\subseteq S_n$,
$d_T(x,X)=d_T(\pi x,\pi X),$
where
$\pi X:=\{\pi x_1,\ldots,\pi x_k\}.$
It follows that
$\pi\M_n(X)=\M_n(\pi X).$
Consequently, for any $x,y\in S_n$,
${\left|\M_{n,k}^{-1}(x)\right|
=
\left|\M_{n,k}^{-1}(y)\right|.}$
Moreover,
$\mu_n(X)=\mu_n(\pi X).$
The same argument applies to the approximate median inverse: since
$\mathcal A_{\pi x}\setminus\cup_i\mathcal A_{\pi x_i}
=\pi\bigl(\mathcal A_x\setminus\cup_i\mathcal A_{x_i}\bigr)$, we have
$\LL_{n,k,c}^{-1}(\pi x)=\pi\LL_{n,k,c}^{-1}(x)$, and hence
$|\LL_{n,k,c}^{-1}(x)|=|\LL_{n,k,c}^{-1}(y)|$ for all $x,y\in S_n$. These probabilities may therefore be analyzed for an arbitrary fixed
$\pi_n\in S_n$, $n\in\N$.

Let $(a_n)_{n\in\mathbbm N}$ be a sequence of nonnegative real numbers,
where $a_n$ is the approximation threshold, and let
$\xi_1^{(n)},\ldots,\xi_k^{(n)}$ be independent uniformly distributed
random permutations in $S_n$. For any sequence $(\pi_n)_{n\in\mathbbm N}$ with $\pi_n\in S_n$, the
uniform distribution of the random $k$-tuple gives
\begin{align*}
\mathbbm P\left(
\pi_n\in
\M_n\bigl(\xi_1^{(n)},\ldots,\xi_k^{(n)}\bigr)
\right)
&=
|\M_{n,k}^{-1}(\pi_n)|/(n!)^k,\\
\mathbbm P\left(
\pi_n\in
\mathscr L_{n,a_n}
\bigl(\xi_1^{(n)},\ldots,\xi_k^{(n)}\bigr)
\right)
&=
|\mathscr L_{n,k,a_n}^{-1}(\pi_n)|/(n!)^k.
\end{align*}

By Theorem~\ref{thmalpha}, on the event
$\{\mathcal O_n(\xi_1^{(n)},\ldots,\xi_k^{(n)})\leq a_n\}$ every exact
median is an $a_n$-approximate median, that is,
$\M_n\subseteq\LL_{n,a_n}$. Consequently,
\begin{equation}\label{eq:restricted}
\begin{aligned}
\mathbbm P\bigl(\pi_n\in\M_n,\ \mathcal O_n\leq a_n\bigr)
&\leq
|\LL_{n,k,a_n}^{-1}(\pi_n)|/(n!)^k,\\
\mathbbm E\bigl[\,|\M_n|\ ;\ \mathcal O_n\leq a_n\,\bigr]
&\leq
|\LL_{n,k,a_n}^{-1}(\pi_n)|/(n!)^{k-1},
\end{aligned}
\end{equation}
where all sets are those of $\xi_1^{(n)},\ldots,\xi_k^{(n)}$. By the
next theorem the event\linebreak $\{\mathcal O_n(\xi_1^{(n)},\ldots,\xi_k^{(n)})\leq a_n\}$ has probability tending to one whenever
$a_n\to\infty$, so counting the approximate median inverse controls the
exact median set outside an event of vanishing probability.
\begin{theorem}\label{thm-inclusion}
Fix $k\geq 2$, and let $(a_n)_{n\in\mathbbm N}$ be a sequence of
positive real numbers such that $a_n\to\infty$. Let
$\xi_1^{(n)},\ldots,\xi_k^{(n)}$ be independent uniformly distributed
random permutations in $S_n$. Then, as $n\to\infty$,
\begin{equation}\label{O_n-conv}
\frac{
\OO_n\bigl(\xi_1^{(n)},\ldots,\xi_k^{(n)}\bigr)
}{a_n}
\longrightarrow 0, \text{ in probability,}
\end{equation}
and
\[
\mathbbm P\left(
\M_n\bigl(\xi_1^{(n)},\ldots,\xi_k^{(n)}\bigr)
\subseteq
\LL_{n,a_n}\bigl(\xi_1^{(n)},\ldots,\xi_k^{(n)}\bigr)
\right)
\longrightarrow 1.
\]
\end{theorem}

\begin{proof}
When $k=2$, we have
$\OO_n\bigl(\xi_1^{(n)},\xi_2^{(n)}\bigr)=0$,
and both conclusions follow immediately. We may therefore assume that
$k\geq 3$.
We first establish a deterministic bound. Let
$X=\{x_1,\ldots,x_\ell\}\subseteq S_n$, $\ell\geq 3$.
From the definition,
we have
$\OO_n(X)\leq \OO_n^{(\ell)}(X).$ On the other hand,
\[
\sum_{1\leq i<j\leq \ell-1}
|\mathcal A_{x_i,x_j}|
=
\sum_{\substack{U\subseteq X\\|U|\geq 2}}
\binom{|U\setminus\{x_\ell\}|}{2}
|\mathcal B_U^X|.
\]
In particular, if
$U\subseteq X\setminus\{x_\ell\}$ and $|U|=r$, then the coefficient of
$|\mathcal B_U^X|$ on the right-hand side is
$\binom{r}{2}\geq r-1.$
It follows that
\[
\OO_n(X)
\leq
\OO_n^{(\ell)}(X)
=
\sum_{r=2}^{\ell-1}(r-1)
\sum_{\substack{
U\subseteq X\setminus\{x_\ell\}\\
|U|=r
}}
|\mathcal B_U^X|
\leq
\sum_{1\leq i<j\leq \ell-1}
|\mathcal A_{x_i,x_j}|.
\]
Consequently, for every realization of
$\xi_1^{(n)},\ldots,\xi_k^{(n)}$,
\[
\OO_n\bigl(\xi_1^{(n)},\ldots,\xi_k^{(n)}\bigr)
\leq
\sum_{1\leq i<j\leq k}
\left|
\mathcal A_{\xi_i^{(n)},\xi_j^{(n)}}
\right|.
\]
This inequality remains valid when repetitions occur in the random sample, since the
left-hand side is computed for the resulting ordinary set by definition, while
repeated sampled permutations only add nonnegative terms to the sum on
the right-hand side. For every $\varepsilon>0$,
\[
\begin{aligned}
\mathbbm P\left(
\OO_n\bigl(\xi_1^{(n)},\ldots,\xi_k^{(n)}\bigr)
\geq \varepsilon a_n
\right)
&\leq
\mathbbm P\left(
\sum_{1\leq i<j\leq k}
|
\mathcal A_{\xi_i^{(n)},\xi_j^{(n)}}
|
\geq \varepsilon a_n
\right)
\\
&\leq
\sum_{1\leq i<j\leq k}
\mathbbm P\left(
|
\mathcal A_{\xi_i^{(n)},\xi_j^{(n)}}
|
\geq
\frac{\varepsilon a_n}{\binom{k}{2}}
\right).
\end{aligned}
\]
By the left-invariance of the breakpoint distance and
Proposition~\ref{tools}, for every fixed $i<j$,
$
|\mathcal A_{\xi_i^{(n)},\xi_j^{(n)}}|/a_n$ converges to zero in probability. Since the number of pairs is fixed, the preceding sum converges to zero.
Therefore, \eqref{O_n-conv} is proved. Finally, as $n\to\infty$, Theorem~\ref{thmalpha} implies 
\begin{multline*}
\mathbbm P\left(
\M_n\bigl(\xi_1^{(n)},\ldots,\xi_k^{(n)}\bigr)
\nsubseteq
\LL_{n,a_n}\bigl(\xi_1^{(n)},\ldots,\xi_k^{(n)}\bigr)
\right)
\\
\qquad\leq
\mathbbm P\left(
\frac{
\OO_n\bigl(\xi_1^{(n)},\ldots,\xi_k^{(n)}\bigr)
}{a_n}
>1
\right)
\longrightarrow 0,
\end{multline*}
which proves the second assertion.
\end{proof}

\begin{remark}
Although Theorem~\ref{thm-inclusion} holds for every sequence
$(a_n)_{n\in\mathbbm N}$ such that $a_n\to\infty$, the meaningful
regime for approximate medians is
$a_n=o(n).$
Indeed, every permutation in $S_n$ has exactly $n-1$ adjacencies, so
if $a_n\geq n-1$, then every permutation is automatically an
$a_n$-approximate median. By contrast, when $a_n=o(n)$, the number of
adjacencies of an approximate median that are absent from all input
permutations is negligible relative to its total number of
adjacencies. Thus, in applications of the theorem, we take
$a_n\to\infty$ sufficiently slowly that $a_n/n\to0$, as $n\to \infty$.
\end{remark}
Theorem~\ref{thm-inclusion} shows that, with high probability, the exact
median set is contained in the $a_n$-approximate median set for any
diverging sequence $(a_n)$. The next theorem, which is our main
motivation for studying approximate medians, shows that approximate
medians are near-optimal and, conversely, that near-optimal permutations
are approximate medians. For $c\geq0$ write
\[
\M_n^{c}(X):=\bigl\{y\in S_n:\ d_T(y,X)\leq\mu_n(X)+c\bigr\}
\]
for the set of $c$-near medians of $X$, so that $\M_n^{0}(X)=\M_n(X)$.

\begin{theorem}[Approximate medians are near-optimal]\label{thm:near-optimal}
Let $X=\{x_1,\ldots,x_k\}\subseteq S_n$, where $x_1,\ldots,x_k$ are
distinct and $k\geq2$. Then, for every $c\geq0$
\begin{enumerate}[(i)]
\item $\LL_{n,c}(X)\subseteq\M_n^{\,c+\Lambda_n(X)}(X)$;
\item $\M_n^{c}(X)\subseteq\LL_{n,\mathcal O_n(X)+c}(X)$;
\item $(k-1)(n-1)-\Lambda_n(X)\leq\mu_n(X)
\leq(k-1)(n-1)-\Lambda_n(X)+\mathcal O_n(X)$.
\end{enumerate}
\end{theorem}
\begin{remark}
Combining (i) and (ii),
\[\LL_{n,c}(X)\subseteq\M_n^{\,c+\Lambda_n(X)}(X)\subseteq
\LL_{n,c+\Lambda_n(X)+\mathcal O_n(X)}(X);\] that is, approximate medians
and near-optimal medians coincide up to a shift of the tolerance by
$\Lambda_n(X)+\mathcal O_n(X)\leq2\Lambda_n(X)$. This follows on
applying (i) and then (ii) with $c+\Lambda_n(X)$ in place of $c$,
together with $\mathcal O_n(X)\leq\Lambda_n(X)$.
\end{remark}

\begin{proof}[Proof of Theorem~\ref{thm:near-optimal}]
Since $\eta_r(y)\leq\sum_{U\subseteq X,\,|U|=r}|\mathcal B_U^X|$ for
every $2\le r\le k$, we have $\sum_{r=2}^k(r-1)\eta_r(y)\leq\Lambda_n(X)$, and
therefore, by~\eqref{eq:dT-eta}, for  any $y\in S_n$,
\begin{equation}\label{eq:dT-sandwich}
(k-1)(n-1)-\Lambda_n(X)+\eta_0(y)\ \leq\ d_T(y,X)\ \leq\ (k-1)(n-1)+\eta_0(y).
\end{equation}
Taking the minimum over $y$ in the left inequality gives the lower bound
in (iii). For the upper bound, the proof of Theorem~\ref{thmalpha} shows
that $\mathcal O_n^{(j)}(X)=\Lambda_n(X)-(k-1)(n-1)+d_T(x_j,X)$, whence
\begin{equation}\label{eq:best-input}
\min_{1\leq j\leq k}d_T(x_j,X)=(k-1)(n-1)-\Lambda_n(X)+\mathcal O_n(X),
\end{equation}
and $\mu_n(X)\leq\min_j d_T(x_j,X)$.

For (i), let $y\in\LL_{n,c}(X)$, so that $\eta_0(y)\leq c$. The right
inequality in~\eqref{eq:dT-sandwich} and the lower bound in (iii) give
$d_T(y,X)\leq(k-1)(n-1)+c\leq\mu_n(X)+\Lambda_n(X)+c$.

For (ii), let $y\in\M_n^{c}(X)$. By~\eqref{eq:dT-eta} and
$\sum_{r=2}^k(r-1)\eta_r(y)\leq\Lambda_n(X)$,
\[\begin{aligned}
    \eta_0(y)&=d_T(y,X)-(k-1)(n-1)+\sum_{r=2}^k(r-1)\eta_r(y)\\
&\leq d_T(y,X)-(k-1)(n-1)+\Lambda_n(X),  
\end{aligned}\]
and $d_T(y,X)\leq\mu_n(X)+c\leq\min_j d_T(x_j,X)+c$, so
that~\eqref{eq:best-input} gives $\eta_0(y)\leq\mathcal O_n(X)+c$.
\end{proof}

\begin{corollary}\label{cor:sandwich}
Fix $k\geq2$, and let $\xi_1^{(n)},\ldots,\xi_k^{(n)}$ be independent
uniformly distributed random permutations in $S_n$. Then
$$\e\,\Lambda_n(\{\xi_1^{(n)},\ldots,\xi_k^{(n)}\})<k(k-1),$$ and for
every integer $b\geq0$, with probability at least $1-k(k-1)/(b+1)$,
\[
\LL_{n,c}\subseteq\M_n^{\,c+b}\subseteq\LL_{n,c+2b}
\qquad\text{for all } c\geq0,
\]
where all sets are those of $\{\xi_1^{(n)},\ldots,\xi_k^{(n)}\}$. In
particular, for random genomes, approximate medians and near-optimal
medians coincide up to a shift of the tolerance that is bounded in
probability.
\end{corollary}

\begin{proof}
Let $X'$ be the set of distinct elements among
$\xi_1^{(n)},\ldots,\xi_k^{(n)}$. If $|X'|=1$, then $\mu_n(X')=0$,
$\M_n^{c}(X')=\LL_{n,c}(X')$ for every $c\geq0$ and $\Lambda_n(X')=0$, so the
inclusions hold trivially. If $|X'|\geq2$, then on the event
$\{\Lambda_n(X')\leq b\}$, Theorem~\ref{thm:near-optimal}(i) and (ii),
together with $\mathcal O_n(X')\leq\Lambda_n(X')$, give
\[
\LL_{n,c}(X')\subseteq\M_n^{\,c+\Lambda_n(X')}(X')\subseteq\M_n^{\,c+b}(X')
\subseteq\LL_{n,\mathcal O_n(X')+c+b}(X')\subseteq\LL_{n,c+2b}(X').
\]
An adjacency occurring in exactly $r$ elements of $X'$ contributes
$r-1$ to $\Lambda_n(X')$ and $\binom r2\geq r-1$ to the sum of
$|\mathcal A_{x,x'}|$ over unordered pairs of distinct $x,x'\in X'$.
The latter sum is at most
$\sum_{1\leq i<j\leq k}|\mathcal A_{\xi_i^{(n)},\xi_j^{(n)}}|$. By
left-invariance and Proposition~\ref{tools}(i),
$\e|\mathcal A_{\xi_i^{(n)},\xi_j^{(n)}}|=2(n-1)/n$ for $i\neq j$, and
therefore
\[
\e\,\Lambda_n(X')\leq\binom k2\frac{2(n-1)}{n}<k(k-1).
\]
Since $\Lambda_n(X')$ is integer-valued, Markov's inequality gives
$\mathbbm P(\Lambda_n(X')>b)=\mathbbm P(\Lambda_n(X')\geq b+1)<k(k-1)/(b+1)$.
\end{proof}

When the input permutations have pairwise maximum breakpoint distance,
Proposition~\ref{extreme} gives $\M_n(X)=\LL_{n,0}(X)$. In general, the
two families appearing in Theorem~\ref{thm:near-optimal} play
complementary roles. The fixed-tolerance sets $\LL_{n,c}(X)$, $c\geq0$,
are monotone in $c$ and, by Theorem~\ref{thm:near-optimal} and
Corollary~\ref{cor:sandwich}, describe the near-optimal solutions at
every level, up to a shift of the tolerance that is bounded in
probability for random inputs. For any diverging tolerance \(a_n\), Theorem~\ref{thm-inclusion} ensures that \(\LL_{n,a_n}\) contains every exact median with probability tending to one.

By left-invariance, $\e|\M_n(\xi_1^{(n)},\ldots,\xi_k^{(n)})|
=n!\,\p(\pi\in\M_n(\xi_1^{(n)},\ldots,\xi_k^{(n)}))$ for every fixed
$\pi\in S_n$, so bounds on this probability translate directly into
bounds on the expected number of exact medians.

For $k=1$ and $n\geq2$, we have
$\mathbbm P\left(
x\in\M_n\bigl(\xi_1^{(n)}\bigr)
\right)
= 2/n!$, for any $x\in S_n.$
For $k=2$, since
$\M_n(y_1,y_2)=\overline{[y_1,y_2]}$,
we obtain
\[
\mathbbm P\left(
x\in\M_n\bigl(\xi_1^{(n)},\xi_2^{(n)}\bigr)
\right)
=
\frac{
\left|
\left\{
(y_1,y_2)\in S_n^2:
x\in\overline{[y_1,y_2]}
\right\}
\right|
}{(n!)^2}\geq
\frac{4}{n!}-\frac{4}{(n!)^2}.
\]
The lower bound counts only the $4n!-4$ ordered pairs for which one
of $y_1$ and $y_2$ is equal to $x$ or its reversal, and therefore
does not include pairs for which $x$ is a non-endpoint element of
$\overline{[y_1,y_2]}$. An explicit upper bound valid for every $k$,
together with the convergence to zero of these probabilities, is given
in Theorem~\ref{thm:unconditional}.

For general $k$, the exact median inverse is considerably more
difficult both to count and to construct: deciding whether a given
permutation is an exact median of given inputs requires solving a
global minimization problem over $S_n$, and finding medians
is NP-hard~\cite{bryant98, sankoff97}. Moreover, numerical experiments
in~\cite{dSJS25b} show that, over the range considered, the average
number of medians of three independent random genomes increases
substantially with $n$. In contrast, whether a permutation is a
$c$-approximate median can be decided directly from its adjacencies and
those of the input genomes, and by Theorem~\ref{thm:near-optimal},
approximate medians are the near-optimal solutions up to a bounded
shift of the tolerance. Their inverse can be constructed explicitly and
admits an explicit counting formula, which we now derive. Exact medians are treated separately, by a factorial-moment argument, in Theorem~\ref{thm:unconditional}.

For every $c\geq0$ and $\pi\in S_n$, the definition of the approximate
median inverse gives
$\mathbbm P(
\pi\in
\LL_{n,c}\bigl(\xi_1^{(n)},\ldots,\xi_k^{(n)}\bigr))
=
\left|\LL_{n,k,c}^{-1}(\pi)\right|/(n!)^k.$
Consequently, by left-invariance,
$$\mathbbm E|
\LL_{n,c}\bigl(\xi_1^{(n)},\ldots,\xi_k^{(n)}\bigr)|
=
\left|\LL_{n,k,c}^{-1}(\pi)\right|/(n!)^{k-1}.$$

The next theorem computes the expected number of $c$-approximate
medians for each fixed $c\geq0$.
To state the result, let
\[
\mathscr P_{k,c}^{(n)}(\pi)
:=
\left\{
(J_1,\ldots,J_k)\in(\mathcal I^{(n)})^k:
J_i\subseteq\mathcal A_\pi \text{ for all }i,
\quad
\left|
\mathcal A_\pi
\setminus
\cup_{i=1}^k J_i
\right|
\leq c
\right\}.
\]
For $c=0$, this agrees with the definition in
Equation~\eqref{index-partition}, so that Theorem~\ref{thm:expected number}
below extends Proposition~\ref{prop:L-inverse-decomposition} to
arbitrary $c$.

\begin{theorem}\label{thm:expected number}
Let $n,k$ be natural numbers, let $c\geq 0$ be a real number, and let $\pi\in S_n$. If $\xi_1^{(n)},\ldots,\xi_k^{(n)}$ are independent uniformly distributed random permutations in $S_n$, then
$$\e|\LL_{n,c}(\xi_1^{(n)},\cdots,\xi_k^{(n)})|=\frac{|\LL_{n,k,c}^{-1}(\pi)|}{(n!)^{k-1}},$$ where
\[
\mathscr{L}_{n,k,c}^{-1}(\pi)=\bigsqcup\limits_{(\tilde{J}_1,...,\tilde{J}_k)\in \mathscr{P}_{k,c}^{(n)}(\pi)} \HH_{\pi}^{(n)}(\tilde J_1)\times ...\times \HH_{\pi}^{(n)}(\tilde J_k),
\]
and
\[
|\mathscr{L}_{n,k,c}^{-1}(\pi)|=\sum\limits_{(\tilde{J}_1,...,\tilde{J}_k)\in \mathscr{P}_{k,c}^{(n)}(\pi)} \prod\limits_{i=1}^k |\HH_{\pi}^{(n)}(\tilde J_i)|.
\]
\end{theorem}
\begin{proof}
From the definition
$$\p(\pi\in \LL_{n,c}(\xi_1^{(n)},\cdots,\xi_k^{(n)}))=|\LL_{n,k,c}^{-1}(\pi)|/(n!)^k.$$
For $x\in S_n$, let the indicator random variable $\delta_x^c=1$ if $x\in \LL_{n,c}(\xi_1^{(n)},\cdots,\xi_k^{(n)})$ and let $\delta_x^c=0$ otherwise. It is then clear that
\[
\begin{split}
\e|\LL_{n,c}(\xi_1^{(n)},\cdots,\xi_k^{(n)})|=\sum\limits_{x\in S_n} \e \delta_x^c= n!|\LL_{n,k,c}^{-1}(\pi)|/(n!)^k.
\end{split}
\]
We must now count the number of elements in $\LL_{n,k,c}^{-1}(\pi)$ in terms of $\mathcal{H}_\pi^{(n)}(J_i)$, for $c>0$. As in the proof of Proposition~\ref{prop:L-inverse-decomposition}, which treats the case $c=0$, let $\mathcal{J}=(J_1,...,J_k),\mathcal{J}'=(J_1',...,J_k')\in \mathscr{P}_{k,c}^{(n)}(\pi)$, such that $\mathcal{J}\neq \mathcal{J}'$, then
\[
(\HH_{\pi}^{(n)}(J_1)\times ...\times \HH_{\pi}^{(n)}(J_k)) \cap (\HH_{\pi}^{(n)}(J_1')\times ...\times \HH_{\pi}^{(n)}(J_k'))=\emptyset.
\]
Now, if $(x_1,...,x_k)\in \mathscr{L}_{n,k,c}^{-1}(\pi)$, then there exist at most $c$ adjacencies of $\pi$ that are not in $\cup_{i=1}^k \mathcal{A}_{x_i}$. Therefore, there exists $(J_1,...,J_k)\in \mathscr{P}_{k,c}^{(n)}(\pi)$ such that, for any $i=1,...,k$, $\mathcal{A}_{\pi,x_i}=J_i$ implying that $(x_1,...,x_k)\in \HH_{\pi}^{(n)}(J_1)\times ...\times \HH_{\pi}^{(n)}(J_k)$. On the other hand, if
\[
(x_1,...,x_k)\in \bigcup\limits_{(\tilde{J}_1,...,\tilde{J}_k)\in \mathscr{P}_{k,c}^{(n)}(\pi)} \HH_{\pi}^{(n)}(\tilde J_1)\times ...\times \HH_{\pi}^{(n)}(\tilde J_k),
\]
then there exists $(J_1,...,J_k)\in \mathscr{P}_{k,c}^{(n)}(\pi)$ such that $x_i\in \HH_{\pi}^{(n)}(J_i)$, for $i=1,...,k$, and so $\mathcal{A}_{\pi,x_i}=J_i$. Thus
\[
\begin{array}{l}
\mathcal{A}_\pi \setminus \bigcup\limits_{i=1}^k J_i=\mathcal{A}_\pi \setminus  \bigcup\limits_{i=1}^k \mathcal{A}_{\pi,x_i} =\mathcal{A}_\pi \setminus \bigcup\limits_{i=1}^k \mathcal{A}_{x_i}.
\end{array}
\]
Therefore, $|\mathcal{A}_\pi \setminus \cup_{i=1}^k \mathcal{A}_{x_i}|\leq c$, and thus, $(x_1,...,x_k)\in \mathscr{L}_{n,k,c}^{-1}(\pi)$.
\end{proof}

\begin{remark}\label{remark:3}
An element $(J_1,\ldots,J_k)\in\mathscr P_{k,c}^{(n)}(\pi)$
can equivalently be represented by a family
$(\tilde J_A)_{A\subseteq[k]}$ of pairwise disjoint subsets satisfying
$\mathcal A_\pi=\bigsqcup_{A\subseteq[k]}\tilde J_A$
and $|\tilde J_\emptyset|\leq c$, where
\[
\tilde J_A
=
(\cap_{i\in A} J_i)
\setminus
(\cup_{j\notin A} J_j), \quad  \text{for} \ \emptyset\neq A \subseteq [k],
\quad \text{and }
\tilde J_\emptyset
=
\mathcal A_\pi\setminus (\cup_{i=1}^kJ_i).
\]
Thus, $\tilde J_A$ consists precisely of those adjacencies of $\pi$
that belong to $J_i$ for every $i\in A$ and to no $J_j$ for
$j\notin A$.
In particular, for $i\in [k]$, $J_i=\bigsqcup_{\substack{i\in A\subseteq[k]}}\tilde J_A.$ Thus, to construct such a family, we first choose the adjacencies
assigned to $\tilde J_\emptyset$ and then assign every remaining
adjacency independently to one of the $2^k-1$ nonempty subsets of
$[k]$. Since every subset of $\mathcal A_\pi$ is a segment set, this
gives a bijection and hence
\[
\left|\mathscr P_{k,c}^{(n)}(\pi)\right|
=
\sum_{j=0}^{\lfloor c\rfloor}
\binom{n-1}{j}(2^k-1)^{n-1-j}.
\]
In particular, for fixed $k\ge 2$ and $c=o(n)$,
\[
|\mathscr P_{k,c}^{(n)}(\pi)|=(2^k-1)^{n+o(n)}.
\]
The term corresponding to $j=0$ gives the lower bound
$(2^k-1)^{n-1}$, while
\[
|\mathscr P_{k,c}^{(n)}(\pi)|
\leq
(2^k-1)^{n-1}
\sum_{j=0}^{\lfloor c\rfloor}\binom{n-1}{j},
\]
and the sum on the right is $\exp(o(n))$ whenever $c=o(n)$.
Thus, allowing a sublinear number of uncovered adjacencies does not
change the exponential growth rate of $\mathscr P_{k,c}^{(n)}(\pi)$.
\end{remark}
By Theorem~\ref{thm:expected number}, computing
$|\mathscr L_{n,k,c}^{-1}(\pi)|$ reduces to computing
$|\HH_\pi^{(n)}(J_i)|$ for the segment sets appearing in
the tuples
$(J_1,\ldots,J_k)\in\mathscr P_{k,c}^{(n)}(\pi)$.
The disjoint-union representation above also describes how a given
permutation becomes a $c$-approximate median. Indeed, for
$(J_1,\ldots,J_k)\in\mathscr P_{k,c}^{(n)}(\pi)$ and any
$x_i\in\HH_\pi^{(n)}(J_i),$
$i\in[k],$
we have
$\mathcal A_{\pi,x_i}=J_i,$
$d(\pi,x_i)=n-1-|J_i|$,
and
\[
|
\mathcal A_\pi\setminus\bigcup_{i=1}^k\mathcal A_{x_i}
|
=
|
\mathcal A_\pi\setminus\bigcup_{i=1}^kJ_i
|
\leq c.
\]
Consequently,
$\pi\in\LL_{n,c}(x_1,\ldots,x_k).$
This observation also gives the following constructive procedure for
generating a tuple having a given permutation $\pi$ as a
$c$-approximate median with prescribed breakpoint distances, whenever
segment sets satisfying the conditions below exist:
\begin{enumerate}
\item Fix or sample a permutation $\pi\in S_n$, and fix integers
$d_1,\ldots,d_k\in\{0,\ldots,n-1\}$ satisfying
$d_1+\cdots+d_k-(k-1)(n-1)\leq c.$

\item Choose, possibly at random, a family
$(\tilde J_A)_{A\subseteq[k]}$ of pairwise disjoint subsets such that
$\mathcal A_\pi=\bigsqcup_{A\subseteq[k]}\tilde J_A$,
$|\tilde J_\emptyset|\leq c$, and, for
$J_i
:=
\cup_{\substack{A\subseteq[k]\\ i\in A}}\tilde J_A,$
$|J_i|=n-1-d_i$
and
$\HH_\pi^{(n)}(J_i)\neq\emptyset,\ i\in[k].$

\item For each $i\in[k]$, choose
$x_i\in\HH_\pi^{(n)}(J_i)$.
Concretely, assign orientations to the segments of $J_i$ and arrange
the oriented segments together with the elements of $[n]$ not
contained in them, subject to the condition that no additional
adjacency of $\pi$ is introduced.
\end{enumerate}

The resulting tuple $(x_1,\ldots,x_k)$ satisfies
$d(\pi,x_i)=n-1-|J_i|=d_i$, $i\in[k],$
and
$\left|
\mathcal A_\pi\setminus
\cup_{i=1}^k\mathcal A_{x_i}
\right|
=
|\tilde J_\emptyset|
\leq c.$ Therefore,
$\pi\in\LL_{n,c}(x_1,\ldots,x_k)$. The procedure above is constructive, with its cost depending on finding a feasible family $(\tilde J_A)_{A\subseteq[k]}$ and constructing elements of the corresponding $\HH_\pi^{(n)}(J_i)$. By Theorem~\ref{thm:near-optimal}(i), the total distance of $\pi$ to $\{x_1,\ldots,x_k\}$ exceeds the median value by at most $c+\Lambda_n(\{x_1,\ldots,x_k\})$. The procedure thus generates sets of genomes for which a near-optimal median with prescribed distances to the input genomes is known in advance; such instances can be used, for example, to test exact and heuristic median algorithms. 

To apply Theorem~\ref{thm:expected number},
it remains to determine
$|\HH_\pi^{(n)}(I)|$ for a segment set
$I\subseteq\mathcal A_\pi$. For a segment set $J$, define
$\mathcal R_n(J)
=
\left\{
x\in S_n:J\subseteq\mathcal A_x
\right\}.$
Since $\HH_\pi^{(n)}(I)$ consists of the permutations that contain
every adjacency of $I$ and none of the adjacencies in
$\mathcal A_\pi\setminus I$, the inclusion--exclusion principle gives
\begin{equation}\label{inclusion-exclusion}
\left|\HH_\pi^{(n)}(I)\right|
=
\sum_{I\subseteq J\subseteq\mathcal A_\pi}
(-1)^{|J\setminus I|}
\left|\mathcal R_n(J)\right|.
\end{equation}

To evaluate this sum, we classify the segment sets
$J$, with $I\subseteq J\subseteq\mathcal A_\pi$, according to their
types. Segment sets of the same type have the same number of
adjacencies and the same number of segments, and hence give the same
value of $|\mathcal R_n(J)|$.

Fix $I\subseteq\mathcal A_\pi$, let $q=\|I\|$, and write
$\mathcal A_\pi\setminus I
=
\bar I_\pi^{(1)}
\sqcup\cdots\sqcup
\bar I_\pi^{(q+1)},$
where the complementary segments are ordered from left to right along
$\pi$. The first and last complementary segments may be empty.

For $I\subseteq J\subseteq\mathcal A_\pi$, the global type of $J$ with
respect to $\pi$ and $I$ is\linebreak
${\lambda
=
(\lambda_1,\ldots,\lambda_{q+1})},$ where $
{\lambda_i
=
\bigl(
\lambda_i^{(1)},
\lambda_i^{(2)},
\lambda_i^{(3)},
\lambda_i^{(4)}
\bigr)}$ is the type associated with the complementary segment
$\bar I_\pi^{(i)}$.
Here,
$\lambda_i^{(1)}
=
|J\cap\bar I_\pi^{(i)}|$
is the number of adjacencies of $J$ contained in
$\bar I_\pi^{(i)}$, and
$\lambda_i^{(2)}
=
\|J\cap\bar I_\pi^{(i)}\|$
is the number of segments of
$J\cap\bar I_\pi^{(i)}$.
Moreover, $\lambda_1^{(3)}=0$, and, for $i=2,\ldots,q+1$, $\lambda_i^{(3)}
=1,$ if the leftmost adjacency of $\bar I_\pi^{(i)}$ belongs to $J$, and $\lambda_i^{(3)}
=0$ otherwise.
Similarly, $\lambda_{q+1}^{(4)}=0$, and, for $i=1,\ldots,q$, $\lambda_i^{(4)}
=1,$ if the rightmost adjacency of $\bar I_\pi^{(i)}$ belongs to $J$, and $\lambda_i^{(4)}
=0$ otherwise.

Set $\|\emptyset\|=0$, and let $\lambda_i=(0,0,0,0)$
when $J\cap\bar I_\pi^{(i)}=\emptyset$.
Put $f_i=\mathbf{1}_{\{i=1\}}+\mathbf{1}_{\{i=q+1\}}$,
the number of ends of $\bar I_\pi^{(i)}$ not adjacent to $I$.
We use $\binom{-1}{-1}=1$ and $\binom uv=0$ unless
$0\le v\le u$ or $u=v=-1$.

The global type determines both the number of adjacencies and the
number of segments of $J$. More precisely,
\[
|J|
=
|I|+\sum_{i=1}^{q+1}\lambda_i^{(1)},
\qquad \text{and} \qquad
\|J\|
=
\|I\|
+
\sum_{i=1}^{q+1}\lambda_i^{(2)}
-
\sum_{i=1}^{q+1}
\bigl(\lambda_i^{(3)}+\lambda_i^{(4)}\bigr).
\]
Indeed, the segments of $J$ are those of $I$ and of the sets $J\cap\bar I_\pi^{(i)}$, except that a segment of $J\cap\bar I_\pi^{(i)}$ containing an end of $\bar I_\pi^{(i)}$ adjacent to $I$ merges with the neighboring segment of $I$; each such merge, counted by $\lambda_i^{(3)}$ or $\lambda_i^{(4)}$, lowers the number of segments by one. Consequently, since
$|\mathcal R_n(J)|
=
2^{\|J\|}(n-|J|)!,$
the value of $|\mathcal R_n(J)|$ depends only on the global type of
$J$ (see Lemma~\ref{permutation}).

For example, let
$\pi=1\,2\,3\,4\,5\,6\,7\,8$
and
$I=
\bigl\{
\{2,3\},\{3,4\},\{5,6\}
\bigr\}.$
Then $\|I\|=2$, and the three complementary segments are
${\bar I_\pi^{(1)}=\{\{1,2\}\}}$, {$\bar I_\pi^{(2)}=\{\{4,5\}\}$}, and
${\bar I_\pi^{(3)}=\{\{6,7\},\{7,8\}\}.}$
Consider
$J
=
I\cup
\bigl\{
\{1,2\},\{4,5\},\{6,7\}
\bigr\}.$
Then
$\lambda
=
\bigl(
(1,1,0,1),
(1,1,1,1),
(1,1,1,0)
\bigr).$
For instance, the second component records that
$J\cap\bar I_\pi^{(2)}$ is one adjacency forming one segment, incident to the segments of \(I\) on both sides.
In this case,
\[
|J|
=
|I|+\sum_{i=1}^3\lambda_i^{(1)}
=6,\quad
\|J\|
=
\|I\|
+\sum_{i=1}^3\lambda_i^{(2)}
-\sum_{i=1}^3
\bigl(\lambda_i^{(3)}+\lambda_i^{(4)}\bigr)
=1.
\]
Therefore,
$|\mathcal R_8(J)|
=
2^{\|J\|}(8-|J|)!
=4.$
Since $|J\setminus I|=3$, this particular segment set contributes
$(-1)^{|J\setminus I|}|\mathcal R_8(J)|
=
-4$
to the inclusion--exclusion sum in
\eqref{inclusion-exclusion}. More generally, all segment sets having the same global type as $J$
give the same value of $|\mathcal R_8(J)|$;
Theorem~\ref{seth} groups the terms in
\eqref{inclusion-exclusion} according to their global types.

\begin{theorem} \label{seth}
Let $\pi$ be a permutation in $S_n$, and $I\in \mathcal{I}^{(n)}$ be a segment set contained in $\pi$. Then
\begin{multline}
|\HH_{\pi}^{(n)}(I)|=\sum\limits_\lambda \left\lbrace (-1)^{\sum\limits_{i=1}^{\|I\|+1} \lambda_i^{(1)}} \prod\limits_{i=1}^{\|I\|+1} \left[  {\lambda_i^{(1)}-1 \choose \lambda_i^{(2)}-1}{|\bar{I}_\pi^{(i)}|-\lambda_i^{(1)}-1+f_i \choose \lambda_i^{(2)} -\lambda_i^{(3)}-\lambda_i^{(4)}} \right]  \times \right. \\
\left. 2^{\{ \|I\|+\sum\limits_{i=1}^{\|I\|+1} \lambda_i^{(2)} - \sum\limits_{i=1}^{\|I\|+1} (\lambda_i^{(3)}+\lambda_i^{(4)})\}}(n-|I|-\sum\limits_{i=1}^{\|I\|+1} \lambda_i^{(1)})! \right\rbrace,
\end{multline}
where the sum is over all
$\lambda=(\lambda_i)_{i=1}^{\|I\|+1}$ such that, for each
$i=1,\ldots,\|I\|+1$, either
$\lambda_i=(0,0,0,0)$, or
\[
1\leq\lambda_i^{(1)}\leq|\bar I_\pi^{(i)}|,
\qquad
1\leq\lambda_i^{(2)}
\leq
\min\left\{
\lambda_i^{(1)},
|\bar I_\pi^{(i)}|+1-\lambda_i^{(1)}
\right\},
\]
with $\lambda_i^{(3)},\lambda_i^{(4)}\in\{0,1\}$.
In addition,
$\lambda_1^{(3)}=\lambda_{\|I\|+1}^{(4)}=0$.

\end{theorem}
To prove Theorem~\ref{seth}, we need the following lemmas.
\begin{lemma}\label{permutation}
Let $I$ be a segment set of $S_n$. Then the number of permutations in $S_n$ containing $I$ is $2^{\|I\|}(n-|I|)!$.
\end{lemma}
\begin{proof}
As the segment set $I$ has $|I|$ adjacencies and $\|I\|$ segments, each permutation containing $I$ has $n-|I|-\|I\|$ points (genes) which are not used in $I$. Noting that segments have two directions, we then have $2^{\|I\|}(\|I\|+(n-|I|-\|I\|))!$ permutations containing $I$.
\end{proof}
\begin{lemma}\label{volume of R}
Let $I,J\in \mathcal{I}^{(n)}$ and $\pi\in S_n$, such that $I\subseteq J\subseteq \mathcal{A}_\pi$. Let $\lambda=(\lambda_i)_{1\leq i\leq \|I\|+1}$, with $\lambda_i=(\lambda_i^{(1)},\lambda_i^{(2)},\lambda_i^{(3)},\lambda_i^{(4)})$, for $i=1,...,\|I\|+1$ be the type of $J$ with respect to $\pi$ and $I$. Then
\[
|\mathcal{R}_n(J)|=2^{\{ \|I\|+\sum\limits_{i=1}^{\|I\|+1} \lambda_i^{(2)} - \sum\limits_{i=1}^{\|I\|+1} (\lambda_i^{(3)}+\lambda_i^{(4)})\}}(n-|I|-\sum\limits_{i=1}^{\|I\|+1} \lambda_i^{(1)})!.
\] 
\end{lemma}
\begin{proof}
By the definition of the type, $J$ has
$|I|+\sum_{i}\lambda_i^{(1)}$ adjacencies and
$\|I\|+\sum_{i}\lambda_i^{(2)}-\sum_{i}(\lambda_i^{(3)}+\lambda_i^{(4)})$
segments, so the claim follows from Lemma~\ref{permutation}.
\end{proof}
\begin{lemma}\label{number of J-type given}
Let $\pi\in S_n$ and $I\subseteq\mathcal A_\pi$. The number of segment sets $J$, with $I\subseteq J\subseteq \mathcal{A}_\pi$ and with type $\lambda=(\lambda_i)_{1\leq i\leq \|I\|+1}$ with respect to $\pi$ and $I$, where $\lambda_i=(\lambda_i^{(1)},\lambda_i^{(2)},\lambda_i^{(3)},\lambda_i^{(4)})$ for $i=1,...,\|I\|+1$, is
\[
\prod\limits_{i=1}^{\|I\|+1} {\lambda_i^{(1)}-1 \choose \lambda_i^{(2)}-1}{|\bar{I}_\pi^{(i)}|-\lambda_i^{(1)}-1+f_i \choose \lambda_i^{(2)} -\lambda_i^{(3)}-\lambda_i^{(4)}}.
\] 
\end{lemma}
\begin{proof}
For each $i=1,\ldots,\|I\|+1$, we count the segment sets
$\tilde J_i=J\cap\bar I_\pi^{(i)}$ having local type $\lambda_i$.
Empty intersections contribute one choice under our binomial
convention, including when $\bar I_\pi^{(i)}=\emptyset$.
We therefore consider nonempty $\tilde J_i$.

Let $\mathbb X_1,\ldots,\mathbb X_{\lambda_i^{(2)}}$ be the
lengths of the selected segments in $\tilde J_i$, and let
$\mathbb Y_1,\ldots,\mathbb Y_{\lambda_i^{(2)}+1}$ be the lengths
of the unselected gaps before, between, and after these segments.
They satisfy the two independent equations
\[
\mathbb X_1+\cdots+\mathbb X_{\lambda_i^{(2)}}
=\lambda_i^{(1)},
\qquad
\mathbb Y_1+\cdots+\mathbb Y_{\lambda_i^{(2)}+1}
=|\bar I_\pi^{(i)}|-\lambda_i^{(1)},
\]
where all $\mathbb X_j$ and all internal gaps
$\mathbb Y_2,\ldots,\mathbb Y_{\lambda_i^{(2)}}$ are positive.

At an end adjacent to $I$, the boundary gap is zero if the
corresponding indicator $\lambda_i^{(3)}$ or $\lambda_i^{(4)}$
equals one, and positive otherwise. The initial gap when $i=1$
and the final gap when $i=\|I\|+1$ may each be zero or positive;
these are precisely the $f_i$ free boundary gaps.

Thus, after omitting the gaps forced to be zero, there remain
$\lambda_i^{(2)}+1-\lambda_i^{(3)}-\lambda_i^{(4)}$ gap variables,
of which $f_i$ are nonnegative and the others positive.
Adding one to each of the $f_i$ nonnegative variables makes all
remaining variables positive and increases their total to
$|\bar I_\pi^{(i)}|-\lambda_i^{(1)}+f_i$.
This gives
\[
{\lambda_i^{(1)}-1\choose\lambda_i^{(2)}-1}
{|\bar I_\pi^{(i)}|-\lambda_i^{(1)}-1+f_i
 \choose\lambda_i^{(2)}-\lambda_i^{(3)}-\lambda_i^{(4)}}
\]
choices for $\tilde J_i$.
If no gap variables remain, the gap equation has one solution
exactly when $\lambda_i^{(1)}=|\bar I_\pi^{(i)}|$, as covered
by our binomial convention.

Finally, the local choices determine
$J=I\cup\bigcup_{i=1}^{\|I\|+1}\tilde J_i$ uniquely.
Multiplying the independent local counts over
$i=1,\ldots,\|I\|+1$ yields the result.
\end{proof}

\begin{proof}[Proof of Theorem~\ref{seth}]
The proof is a direct application of Lemmas \ref{volume of R} and \ref{number of J-type given}, and (\ref{inclusion-exclusion}), the inclusion-exclusion principle.
\end{proof}

The global type representation in Theorem~\ref{seth} already reduces
the number of terms relative to direct inclusion--exclusion.
A further reduction is possible because distinct global types may
produce the same values of $|J\setminus I|$ and
$\|J\|$. Using dynamic programming, we can group all such types and evaluate their combined contribution without enumerating every global type separately.

Let
$q=\|I\|,$
$m=n-1-|I|,$
and, for each $i\in[q+1]$, let $\mathcal T_i$ denote the set of
admissible local types
$\lambda_i
=
\bigl(
\lambda_i^{(1)},
\lambda_i^{(2)},
\lambda_i^{(3)},
\lambda_i^{(4)}
\bigr)$
in the complementary segment $\bar I_\pi^{(i)}$. Let
$w_i(\lambda_i)$ denote the number of subsets of
$\bar I_\pi^{(i)}$ having local type $\lambda_i$, as given by Lemma~\ref{number of J-type given}.
In particular, $(0,0,0,0)\in\mathcal T_i$ and
$w_i(0,0,0,0)=1$.
For a global type $\lambda$, define
\[
a(\lambda)
=
\sum_{i=1}^{q+1}\lambda_i^{(1)},
\qquad
b(\lambda)
=
q+
\sum_{i=1}^{q+1}
\left(
\lambda_i^{(2)}
-\lambda_i^{(3)}
-\lambda_i^{(4)}
\right).
\]
By the identities preceding Theorem~\ref{seth}, if $J$ has global
type $\lambda=(\lambda_1,\ldots,\lambda_{q+1})$, then
$a(\lambda)=|J\setminus I|$ and $b(\lambda)=\|J\|$.

Consequently, the formula in Theorem~\ref{seth} may be written as
\[
\left|\HH_\pi^{(n)}(I)\right|
=
\sum_{\lambda_1\in\mathcal T_1}
\cdots
\sum_{\lambda_{q+1}\in\mathcal T_{q+1}}
\left(
\prod_{i=1}^{q+1}w_i(\lambda_i)
\right)
(-1)^{a(\lambda)}
2^{b(\lambda)}
\bigl(n-|I|-a(\lambda)\bigr)!.
\]
Thus, apart from its multiplicity
$\prod_{i=1}^{q+1}w_i(\lambda_i),$
the contribution of a global type depends on it only through the pair
$\bigl(a(\lambda),b(\lambda)\bigr).$
Global types producing the same pair can therefore be combined by
dynamic programming.

For a partial choice of local types
$(\lambda_1,\ldots,\lambda_i)$, define
\[
a_i=
\sum_{j=1}^{i}\lambda_j^{(1)}, \quad
b_i=
q+
\sum_{j=1}^{i}
(
\lambda_j^{(2)}
-\lambda_j^{(3)}
-\lambda_j^{(4)}
).
\]
Note that $a_{q+1}=a(\lambda)$ and $b_{q+1}=b(\lambda).$ For any corresponding local choices $\tilde J_j$ of types
$\lambda_j$, the partial set
$J^{(i)}:=I\cup\bigcup_{j=1}^{i}\tilde J_j$ satisfies
$|J^{(i)}\setminus I|=a_i$ and $\|J^{(i)}\|=b_i$.
In particular, $0\leq a_i\leq m$ and $0\leq b_i\leq n-1$. For $i=0,\ldots,q+1$, let $D_i(r,s)$ be the total multiplicity of
the partial choices
$(\lambda_1,\ldots,\lambda_i)
\in
\mathcal T_1\times\cdots\times\mathcal T_i$ for which $a_i=r$ and $b_i=s.$
More precisely,
\[
D_i(r,s)
=
\sum_{\substack{
(\lambda_1,\ldots,\lambda_i)
\in
\mathcal T_1\times\cdots\times\mathcal T_i\\
a_i=r, \ b_i=s
}}
\prod_{j=1}^{i}w_j(\lambda_j).
\]
For $i=0$, the empty product is equal to one, hence $D_0(0,q)=1$ and $D_0(r,s)=0$ for $(r,s)\neq (0,q)$.
For $i=1,\ldots,q+1$, the quantities $D_i(r,s)$ satisfy the
recurrence
\[
D_i(r,s)
=
\sum_{\lambda_i\in\mathcal T_i}
w_i(\lambda_i)
D_{i-1}
\left(
r-\lambda_i^{(1)},
s-\lambda_i^{(2)}
+\lambda_i^{(3)}
+\lambda_i^{(4)}
\right),
\]
where values with inadmissible indices are interpreted as zero. Indeed, appending a
local type $\lambda_i$ increases the number of added adjacencies by
$\lambda_i^{(1)}$ and changes the accumulated number of segments by
$\lambda_i^{(2)}
-\lambda_i^{(3)}
-\lambda_i^{(4)}.$
After all complementary segments have been processed,
\[
\left|\HH_\pi^{(n)}(I)\right|
=
\sum_{r=0}^{m}
\sum_{s=0}^{n-1}
(-1)^r
D_{q+1}(r,s)
2^s
(n-|I|-r)!.
\]
Accordingly, after processing the first $i$ complementary segments,
the dynamic program combines all partial choices producing the same
pair $(r,s)$ into the single value $D_i(r,s)$, which records their
accumulated multiplicity.

The computational costs of the preceding exact evaluations can be summarized as follows.
Direct inclusion--exclusion enumerates $2^m$ segment sets, where
$m=n-1-|I|$. For the type-based formula, let
$N_i=|\mathcal T_i|$. Direct evaluation enumerates
$\prod_{i=1}^{q+1}N_i$ global types and requires
$O\left((q+1)\prod_{i=1}^{q+1}N_i\right)$
arithmetic operations. If
$s_i=|\bar I_\pi^{(i)}|$, then
$N_i=O((s_i+1)^2)$, and hence the cost is at most
$O\left(
(q+1)\prod_{i=1}^{q+1}(s_i+1)^2
\right).$
Thus it is polynomial in $n$ for fixed $q=\|I\|$, but not uniformly
polynomial when $q$ varies with $n$.
The dynamic program avoids this product. Each $D_i$ has at most
$(m+1)n$ entries, and its update considers the $N_i$ admissible local
types. Hence the total cost is
$O\left(
(m+1)n\sum_{i=1}^{q+1}N_i
\right)
=
O\left(
(m+1)n\sum_{i=1}^{q+1}(s_i+1)^2
\right)$
arithmetic operations, with $O((m+1)n)$ memory when only two consecutive
tables are retained. Since
$\sum_{i=1}^{q+1}s_i=m\leq n-1$, this gives the uniform bound
$O(n^4)$ on the number of arithmetic operations.

For the computational comparison, we take
$\pi=1\,2\,\cdots\,n$, $|I|=8$, and $\|I\|=4$, with each segment of
$I$ containing two adjacencies. For each $n$, we generate $50$ random
configurations of the $m=n-1-|I|$ complementary adjacencies among the
five complementary segments, requiring the three internal segments to
be nonempty. In the direct type-based evaluation, the number of global types is the
number of summands evaluated in the reformulation of
$|\HH_\pi^{(n)}(I)|$ above. In the dynamic program (DP), a DP update is one
application of the recurrence contributing a local type to a state
$D_i(r,s)$. Thus, the DP replaces the enumeration of global types by
these state updates, after which $|\HH_\pi^{(n)}(I)|$ is obtained from
the final table $D_{q+1}$. The reported numbers of global types, DP updates, and
running times are averaged over these configurations. For each
configuration, the type formula is timed in three runs and the dynamic
program in three batches of $100$ evaluations. Since direct
inclusion--exclusion always has $2^m$ terms, independently of the
configuration, one reference timing is reported for each $n$.

\begin{table}[ht!]
\centering
\caption{Comparison of three exact methods for computing
$|\HH_\pi^{(n)}(I)|$ on random segment-set configurations.}
\vspace*{1mm}
\label{tab:theorem4-computation}
\resizebox{\textwidth}{!}{
\begin{tabular}{c c@{~~~~} c c c c c c}
\toprule
$n$
&
$m$
&
\begin{tabular}{c}
Incl--excl\\
terms
\end{tabular}
&
\begin{tabular}{c}
Global types in\\
Theorem~\ref{seth}
\end{tabular}
&
\begin{tabular}{c}
DP\\
updates
\end{tabular}
&
\begin{tabular}{c}
Incl--excl\\
time (s)
\end{tabular}
&
\begin{tabular}{c}
Theorem~\ref{seth}\\
time (s)
\end{tabular}
&
\begin{tabular}{c}
DP\\
time (ms)
\end{tabular}
\\
\midrule
21 & 12 &       4,096 &       2,071 &   422 &   0.0017 & 0.00136 & 0.191 \\
25 & 16 &      65,536 &      15,335 & 1,097 &   0.0266 & 0.00976 & 0.387 \\
29 & 20 &   1,048,576 &      89,418 & 2,362 &   0.4347 & 0.05867 & 0.784 \\
33 & 24 &  16,777,216 &     365,167 & 4,501 &   7.9661 & 0.24700 & 1.359 \\
35 & 26 &  67,108,864 &     755,181 & 5,886 &  33.2816 & 0.52895 & 1.805 \\
37 & 28 & 268,435,456 &     885,998 & 7,891 & 133.1820 & 0.63079 & 2.382 \\
\bottomrule
\end{tabular}
}
\end{table}

Table~\ref{tab:theorem4-computation} illustrates the successive
reductions obtained from the global-type representation and the
dynamic program. Direct inclusion--exclusion exceeds $268$ million
terms for the largest instance. For $n=37$, the
Theorem~\ref{seth} formula involves approximately $886{,}000$ global
types on average, while the dynamic program requires approximately
$7{,}891$ updates. The corresponding average running times are about
$0.63$ seconds and $2.4$ milliseconds, respectively. Thus, the type
representation already provides a substantial computational reduction
over direct inclusion--exclusion, while the dynamic program gives a
further and much larger improvement. 

We finally consider the exact evaluation of the full approximate-median
inverse. Although Theorem~\ref{thm:expected number} expresses
$|\mathscr L_{n,k,c}^{-1}(\pi)|$ as a sum over
$\mathscr P_{k,c}^{(n)}(\pi)$, direct evaluation quickly becomes
impractical. For example,
$|\mathscr P_{3,1}^{(10)}(\pi)|=92{,}236{,}816$.
The same sum can instead be evaluated by subset transforms without
enumerating its terms.

For $E\subseteq\mathcal A_\pi$, let
$F(E)
=
\left|
\left\{
x\in S_n:
E\cap\mathcal A_x=\emptyset
\right\}
\right|.$
By inclusion--exclusion,
\[
F(E)
=
\sum_{T\subseteq E}
(-1)^{|T|}
2^{\|T\|}(n-|T|)!.
\]
Thus all values $F(E)$ can be computed simultaneously by a subset
zeta transform. Moreover, $F(E)^k$ counts the ordered $k$-tuples for
which every adjacency in $E$ is absent from all input permutations. Let $G(M)$ count the ordered $k$-tuples whose set of missing
adjacencies in $\mathcal A_\pi$ is exactly $M$.
Since $F(E)^k=\sum_{E\subseteq M\subseteq\mathcal A_\pi}G(M)$,
superset M\"obius inversion gives
\[
G(M)=
\sum_{M\subseteq E\subseteq\mathcal A_\pi}
(-1)^{|E|-|M|}F(E)^k.
\]
Consequently,
\[
|\mathscr L_{n,k,c}^{-1}(\pi)|
=
\sum_{\substack{M\subseteq\mathcal A_\pi\\
|M|\leq\lfloor c\rfloor}}G(M).
\]
For fixed $k$, this requires $O(n2^n)$ arithmetic operations and
$O(2^n)$ memory.

\begin{table}[ht!]
\centering
\caption{Exact computation of the approximate-median inverse for
$k=3$ and $c=\lfloor n/10\rfloor$.  
}
\footnotesize\begin{tabular*}{\textwidth}{@{~}@{\extracolsep{\fill}}ccccc@{~}}
\toprule
$n$ & $c$
& $|\mathscr L_{n,3,c}^{-1}(\pi)|$
& $\mathbb P(\pi\in\mathscr L_{n,c})$
& Time (s) \\
\midrule
10 & 1 & $7.61\times10^{17}$ & $1.59\times10^{-2}$ & 0.0005 \\
14 & 1 & $3.66\times10^{28}$ & $5.52\times10^{-5}$ & 0.0093 \\
18 & 1 & $1.20\times10^{40}$ & $4.56\times10^{-8}$ & 0.184 \\
22 & 2 & $5.61\times10^{53}$ & $3.95\times10^{-10}$ & 3.63 \\
24 & 2 & $1.40\times10^{60}$ & $5.87\times10^{-12}$ & 15.40 \\
26 & 2 & $4.65\times10^{66}$ & $7.09\times10^{-14}$ & 66.44 \\
28 & 2 & $2.02\times10^{73}$ & $7.11\times10^{-16}$ & 309.64 \\
\bottomrule
\end{tabular*}
\label{tab:inverse-computation}
\end{table}

By left-invariance, we may take $\pi=id$. Table~\ref{tab:inverse-computation}
reports the computation for $k=3$ and $c=\lfloor n/10\rfloor$.
For independent uniform random permutations
$\xi_1^{(n)},\xi_2^{(n)},\xi_3^{(n)}$, the probability column is
$\mathbbm P\left(
\pi\in
\mathscr L_{n,c}
(\xi_1^{(n)},\xi_2^{(n)},\xi_3^{(n)})
\right)
=|\mathscr L_{n,3,c}^{-1}(\pi)|/(n!)^3.$
The probability that a fixed permutation is a $c$-approximate median
decreases sharply over the range considered. In contrast, since
$\mathbbm E|\mathscr L_{n,c}|$ $
=
n!\,\mathbbm P(\pi\in\mathscr L_{n,c}),$
the expected number of approximate medians increases from approximately
$5.8\times10^4$ for $n=10$ to $2.17\times10^{14}$ for $n=28$.
The exact inverse count for $n=28$ is obtained in approximately
$5.2$ minutes. Computations reported above were performed using
Python~3.11.4 on a computer with a $2.3$~GHz quad-core Intel Core~i7
processor and $32$~GB of RAM.

The computations above give exact membership probabilities and
expected cardinalities for $\LL_{n,c}$. We next give explicit
unconditional upper bounds for these quantities for both
$\LL_{n,c}$ and $\M_n^{c}$. Since $\M_n^{0}=\M_n$, this also
bounds the corresponding quantities for $\M_n$ without the
event restriction in~\eqref{eq:restricted}.
Theorem~\ref{thm:abundance} then compares $|\M_n|$ with
$|\LL_{n,\mathcal O_n+s}|$.
\begin{theorem}\label{thm:unconditional}
For integers $k\geq2$ and $c\geq0$, let $\xi_1^{(n)},\ldots,\xi_k^{(n)}$ be independent
uniformly distributed random permutations in $S_n$, and let
$\zeta_1^{(n)},\ldots,\zeta_k^{(n)}$ be independent copies of
$\zeta^{(n)}:=|\mathcal A_{id,\xi^{(n)}}|$, where $\xi^{(n)}$ is
uniformly distributed on $S_n$. Then, for every $\pi\in S_n$,
\begin{equation}\label{upperboundexactmed}
\p\bigl(\pi\in\M_n^{c}(\xi_1^{(n)},\ldots,\xi_k^{(n)})\bigr)
\leq
\p\Bigl(\sum_{i=1}^k\zeta_i^{(n)}\geq n-1-c\Bigr)
\leq
\frac{(2k)!\,(2k)^{n-1-c-2k}}{(n-1-c)!},
\end{equation}
the second inequality holding for $n\geq2k+c+1$. For $n\geq 2k+c+1$,
\[
\e\bigl|\M_n^{c}(\xi_1^{(n)},\ldots,\xi_k^{(n)})\bigr|
\leq
n^{c+1}(2k)!\,(2k)^{n-1-c-2k},
\]
and in particular, for fixed $c$,
$\p\bigl(\pi_n\in\M_n^{c}(\xi_1^{(n)},\ldots,\xi_k^{(n)})\bigr)\to0$ as
$n\to\infty$, for every sequence $(\pi_n)_{n\in\N}$ with $\pi_n\in S_n$.
The same bounds hold with $\M_n^{c}$ replaced by $\LL_{n,c}$.
\end{theorem}
\begin{proof}
Let $X'$ be the set of distinct elements among
$\xi_1^{(n)},\ldots,\xi_k^{(n)}$ and let $k'=|X'|$. If
$y\in\M_n^{c}(X')$, then, for any $x'\in X'$,
\[
k'(n-1)-\sum_{x\in X'}|\mathcal A_{y,x}|
=d_T(y,X')\leq\mu_n(X')+c\leq d_T(x',X')+c\leq(k'-1)(n-1)+c,
\]
so that
$\sum_{i=1}^k|\mathcal A_{y,\xi_i^{(n)}}|
\geq\sum_{x\in X'}|\mathcal A_{y,x}|\geq n-1-c$. If
$y\in\LL_{n,c}(X')$, then also
$\sum_{i=1}^k|\mathcal A_{y,\xi_i^{(n)}}|\geq|\mathcal A_y\cap\cup_i\mathcal A_{\xi_i^{(n)}}|\geq n-1-c$.
Hence both $\{\pi\in\M_n^{c}(\xi_1^{(n)},\ldots,\xi_k^{(n)})\}$ and
$\{\pi\in\LL_{n,c}(\xi_1^{(n)},\ldots,\xi_k^{(n)})\}$ are contained in
$\bigl\{\sum_{i=1}^k|\mathcal A_{\pi,\xi_i^{(n)}}|\geq n-1-c\bigr\}$.
By left-invariance,
$|\mathcal A_{\pi,\xi_i^{(n)}}|=|\mathcal A_{id,\pi^{-1}\xi_i^{(n)}}|$,
and $\pi^{-1}\xi_1^{(n)},\ldots,\pi^{-1}\xi_k^{(n)}$
are independent and uniformly distributed, so the
$|\mathcal A_{\pi,\xi_i^{(n)}}|$ are independent copies of
$\zeta^{(n)}$. This gives the first inequality
in~\eqref{upperboundexactmed}.

For the closed form, note that $\binom{\zeta^{(n)}}{m}$ counts the sets
$T\subseteq\mathcal A_{id}$ with $|T|=m$ and
$T\subseteq\mathcal A_{\xi^{(n)}}$. By Lemma~\ref{permutation} and
$\|T\|\leq|T|$, for $0\leq m\leq n-1$,
\[
\e\binom{\zeta^{(n)}}{m}
=\sum_{\substack{T\subseteq\mathcal A_{id}\\ |T|=m}}
\frac{2^{\|T\|}(n-m)!}{n!}
\leq2^m\binom{n-1}{m}\frac{(n-m)!}{n!}
=\frac{2^m}{m!}\cdot\frac{n-m}{n}\leq\frac{2^m}{m!}.
\]
Let $N=\sum_{i=1}^k\zeta_i^{(n)}$. By Vandermonde's identity,
independence and the multinomial theorem,
\[
\e\binom Nm
=\sum_{m_1+\cdots+m_k=m}\ \prod_{i=1}^k\e\binom{\zeta_i^{(n)}}{m_i}
\leq\sum_{m_1+\cdots+m_k=m}\ \prod_{i=1}^k\frac{2^{m_i}}{m_i!}
=\frac{(2k)^m}{m!}.
\]
If $N\geq t\geq m\geq0$ are integers, then
$\binom Nm\geq\binom tm>0$, so
$\mathbf 1\{N\geq t\}\leq\binom Nm/\binom tm$. Taking expectations with
$t=n-1-c$ and $m=n-1-c-2k$ gives
\[
\p(N\geq n-1-c)
\leq\frac{(2k)^{m}}{m!}\cdot\frac{m!\,(2k)!}{(n-1-c)!}
=\frac{(2k)!\,(2k)^{n-1-c-2k}}{(n-1-c)!},
\]
which is the second inequality in~\eqref{upperboundexactmed}.

Finally, since $\mu_n(\pi X)=\mu_n(X)$ and $d_T(\pi y,\pi X)=d_T(y,X)$, left-invariance gives
$\e|\M_n^{c}(\xi_1^{(n)},\ldots,\xi_k^{(n)})|=n!\,\p(\pi\in\M_n^{c}(\xi_1^{(n)},\ldots,\xi_k^{(n)}))$
for every fixed $\pi\in S_n$, and similarly for $\LL_{n,c}$. Together
with~\eqref{upperboundexactmed} and $n!/(n-1-c)!\leq n^{c+1}$, this gives
the bounds on the expectations, and
$(2k)!\,(2k)^{n-1-c-2k}/(n-1-c)!\to0$ as $n\to\infty$ for fixed $c$.
\end{proof}
In particular, for fixed $c$, $\e|\M_n^{c}(\xi_1^{(n)},\ldots,\xi_k^{(n)})|$ and $\e|\LL_{n,c}(\xi_1^{(n)},\ldots,\xi_k^{(n)})|$ grow at most exponentially in $n$. These bounds are explicit but crude: for $k=3$, the bound on $\e|\LL_{n,c}|$ exceeds the exact values obtained from Table~\ref{tab:inverse-computation} by a factor growing from about $45$ at $n=10$ to about $4\times10^{7}$ at $n=28$.

The sets $\LL_{n,c}$ grow rapidly with the tolerance: for $n=22$ and $k=3$, $\e|\LL_{n,c}|\approx2.3\times10^{8}$, $1.5\times10^{10}$ and $4.4\times10^{11}$ for $c=0,1,2$. The proof of the next theorem shows that, when $k,c=o(n)$,
increasing the tolerance by one increases the cardinality
by a factor of at least $(1-o(1))n/(c+1)$. For fixed $k$,
the theorem then uses $\M_n\subseteq\LL_{n,\mathcal O_n}$
to quantify how much smaller $\M_n$ is than
$\LL_{n,\mathcal O_n+s}$.
\begin{theorem}[Exact versus approximate median counts]\label{thm:abundance}
For fixed $k\ge 2$, let
$\mathcal X_n=\{\xi_1^{(n)},\ldots,\xi_k^{(n)}\}$ be a set formed by $k$ independent uniformly distributed random permutations in $S_n$, and write
$\mathcal O_n=\mathcal O_n(\mathcal X_n)$. Then
\[
\frac{|\M_n(\mathcal X_n)|}{|\LL_{n,\mathcal O_n+s}(\mathcal X_n)|}
=O_{\p}(n^{-s})
\qquad\text{for every fixed integer } s\geq1,
\]
and $n^{p}|\M_n(\mathcal X_n)|/|\LL_{n,a_n}(\mathcal X_n)|\to0$ in
probability for every fixed $p\geq0$ and every $a_n\to\infty$.
\end{theorem}

\begin{proof}
Let $X\subseteq S_n$ be deterministic with $|X|=k\geq2$, write
$\alpha_r=|\{y\in S_n:\eta_0(y)=r\}|$ and abbreviate
$\gamma_c=|\LL_{n,c}(X)|=\sum_{r=0}^{c}\alpha_r$. We first claim that
$(r+1)\alpha_{r+1}\geq(n-1-2k-r)\alpha_r$ for $0\leq r\leq n-2$.
Reversing the prefix $y_1\cdots y_i$, with $1\leq i<n$, replaces only
$\{y_i,y_{i+1}\}$ by $\{y_1,y_{i+1}\}$, so it raises $\eta_0$ by exactly
one unless it removes an adjacency absent from all inputs, which happens
for at most $\eta_0(y)=r$ values of $i$, or creates one present in an
input, which happens for at most $2k$ values, since
$\cup_{x\in X}\mathcal A_x$ has maximum degree at most $2k$. Hence at
least $n-1-2k-r$ of the $n-1$ reversals send $y$ into $\{\eta_0=r+1\}$.
Conversely, each such image has at most $r+1$ preimages, since the
reversal must be undone at one of its $r+1$ adjacencies absent from all
inputs, and each of these determines the preimage uniquely. Double
counting the pairs $(y,i)$ proves the claim, which is immediate when its
right-hand side is negative.

Summing the claim over $r=0,\ldots,c$ and using
$\sum_{r=0}^{c}r\alpha_r\leq c\,\gamma_c$ gives
$(c+1)\alpha_{c+1}\geq(n-1-2k-2c)\gamma_c$, hence
$\gamma_{c+1}/\gamma_c\geq(n-2k-c)/(c+1)$ for $0\leq c\leq n-2$.
Applying this successively at $c,c+1,\ldots,c+s-1$ gives, for all
integers $c,s\geq0$ with $c+s\leq n-2k$,
\begin{equation}\label{LLproportion}
\frac{|\LL_{n,c+s}(X)|}{|\LL_{n,c}(X)|}
=\frac{\gamma_{c+s}}{\gamma_c}
\geq\prod_{j=0}^{s-1}\frac{n-2k-c-j}{c+j+1}
=\frac{\binom{n-2k-c}{s}}{\binom{c+s}{s}}.
\end{equation}
Since $\M_n(X)\subseteq\LL_{n,\mathcal O_n(X)}(X)$ by
Theorem~\ref{thm:near-optimal}(ii) applied with tolerance $0$, taking
$c=\mathcal O_n(X)$ in (\ref{LLproportion}) gives, whenever
$\mathcal O_n(X)+s\leq n-2k$,
\begin{equation}\label{LMproportion}
\frac{|\LL_{n,\mathcal O_n(X)+s}(X)|}{|\M_n(X)|}
\geq
\frac{\binom{n-2k-\mathcal O_n(X)}{s}}{\binom{\mathcal O_n(X)+s}{s}}.
\end{equation}

Now let $X=\mathcal X_n$. The bounds above remain valid for fewer than
$k$ distinct inputs, using the same degree bound $2k$; the singleton
case is immediate. The right-hand side of (\ref{LMproportion}) is nonincreasing in
$\mathcal O_n(X)$, for all sufficiently large $n$, so on
$\{\mathcal O_n\leq K\}$ the ratio in the statement is at most
$\binom{K+s}{s}/\binom{n-2k-K}{s}=O(n^{-s})$ for each fixed $K$; since $(\mathcal O_n)_n$ is tight, by $\mathcal O_n\leq\Lambda_n$ and Corollary~\ref{cor:sandwich}, $K$ may be
chosen with $\p(\mathcal O_n\leq K)$ arbitrarily close to~$1$. Given
$p\geq0$ and $a_n\to\infty$, choose an integer $s>p$; then
$\p(a_n\geq\mathcal O_n+s)\to1$, and the last assertion follows from the
first together with the monotonicity of $c\mapsto|\LL_{n,c}(X)|$.
\end{proof}

For fixed $k$ and $c$, the proof shows that each additional unit of
tolerance increases the cardinality by at least a factor of order $n$,
while increasing the near-optimality bound of
Theorem~\ref{thm:near-optimal}(i) by one. Although the exact and
approximate median sets coincide at zero tolerance when the inputs are
at pairwise maximum breakpoint distance
(Proposition~\ref{extreme}), Theorem~\ref{thm:abundance} shows that the
median proportion decays at least polynomially in probability when the
tolerance exceeds $\mathcal O_n$ by a fixed positive integer.
The precise asymptotic order of the expected number of exact
medians remains open.

\section{Discussion}\label{sec:conclusion}

We introduced the notions of median inverse and approximate median
inverse and used them to study the probability that any given
permutation $x\in S_n$ is a $c$-approximate median of $k$ permutations
$\mathcal X=\{\xi_1^{(n)},\ldots,\xi_k^{(n)}\}$ chosen uniformly and
independently at random from $S_n$. Due to the left-invariance of the
breakpoint distance, this probability is the same for all permutations
in $S_n$. Consequently, we computed the expected number of
$c$-approximate medians of $\mathcal X$. The key was to observe that
any exact median of $\mathcal X$ can have at most
$\mathcal O_n(\mathcal X)$ adjacencies not belonging to
$\bigcup_{i=1}^k\mathcal A_{\xi_i^{(n)}}$. For random inputs, we showed
that
$\mathcal O_n(\mathcal X)/a_n\to 0$
in probability as $n\to\infty$, for any diverging sequence
$(a_n)_{n\in\mathbb N}$. In particular, this is informative when $a_n\to\infty$ with
$a_n=o(n)$.

To determine the probability of a permutation $x$ being a
$c$-approximate median of $\mathcal X$, we count the approximate median inverse
$\mathscr L_{n,k,c}^{-1}(x)$. Splitting $\mathcal A_x$ into the sets
$\tilde J_A$ of adjacencies of $x$ present exactly in the inputs indexed
by $A\subseteq[k]$ reduces this count to the numbers $|\HH_x^{(n)}(J_i)|$
of permutations sharing exactly $J_i=\cup_{A\ni i}\tilde J_A$ with $x$,
which Theorem~\ref{seth} gives explicitly and
Theorem~\ref{thm:expected number} combines. As discussed after
Remark~\ref{remark:3}, the same decomposition gives a constructive
procedure for generating input genomes that have a given permutation
as a $c$-approximate median at prescribed breakpoint distances, whenever
the required segment sets exist.

Our results provide two complementary justifications for considering
the $c$-approximate median set as a relaxation of the exact median set.
First, Theorem~\ref{thm-inclusion} shows that, for random inputs, the exact
median set is contained with high probability in an approximate median
set with a suitably growing tolerance. Second,
Theorem~\ref{thm:near-optimal} shows that approximate medians are
precisely the near-optimal solutions, up to a shift of the tolerance by
$\Lambda_n(X)+\mathcal O_n(X)$, which by Corollary~\ref{cor:sandwich} is
bounded in probability for random genomes. Taking the tolerance to be
zero, the same theorem also contains the exact median set in
$\LL_{n,\mathcal O_n(X)}(X)$ for every set of input genomes. Separately, Theorem~\ref{thm:unconditional} gives explicit upper bounds, with no restriction to a favorable event, on the probability that a given permutation belongs to $\M_n^{c}$ and on $\e|\M_n^{c}|$; the case $c=0$ concerns exact medians, and the same bounds hold for $\LL_{n,c}$. Theorem~\ref{thm:abundance} quantifies the gap between the exact and approximate median sets: for fixed \(k\) and \(c\), each additional unit of tolerance increases the number of approximate medians by at least a factor proportional to \(n\). Moreover, for every fixed integer \(s\geq1\), the proportion of exact medians in \(\LL_{n,\mathcal O_n+s}\) is \(O_{\mathbb P}(n^{-s})\), while, by Remark~\ref{rem:attained}, the tolerance $\mathcal O_n$ in the inclusion $\M_n\subseteq\LL_{n,\mathcal O_n}$ cannot be lowered in general.

In summary, our analysis computes the probability that a given
permutation is a $c$-approximate median of random genomes and the
expected number of such approximate medians, provides a systematic procedure for
generating them, and yields upper bounds for the corresponding
quantities for exact medians. Dynamic programming and subset transforms make these approximate-median quantities exactly computable in practice (Tables~\ref{tab:theorem4-computation} and~\ref{tab:inverse-computation}). 
Our findings provide a foundation for
algorithmic approaches to counting and constructing median and
approximate median genomes.
\bibliographystyle{acm} 
\bibliography{biblio}

\end{document}